\documentclass[footinbib,reprint,aps,pra,superscriptaddress]{revtex4-1}

\usepackage{amsmath}  
\usepackage{graphicx}
\usepackage[usenames]{xcolor}
\usepackage{caption}
\usepackage{subcaption}
\DeclareGraphicsExtensions{.pdf}
\graphicspath{ {./images/} }
\usepackage{hyperref}
\usepackage{physics}
\usepackage{color}

\newcommand{\expec}[1]{\langle #1 \rangle}
\newcommand{\matr}[1]{\mathbf{#1}}
\newcommand{\nbarTm}[0]{\bar{n}_m^T}
\newcommand{\nbarTc}[0]{\bar{n}_c^T}
\newcommand{\lambdamax}[0]{\lambda_{\mathrm{max}}}
\newcommand{\nTeffout}[0]{n^T_{\mathrm{eff, out}}}
\newcommand{\gain}[0]{\mathcal{G}}
\newcommand{\nAddAmp}[0]{\bar{n}_{\mathrm{add}}^{(\mathrm{amp})}}
\newcommand{\nAddFD}[0]{\bar{n}_{\mathrm{add}}^{(\mathrm{FD})}}

\begin{document}

\title{Optomechanics with two-phonon driving}

\author{B.~A.~Levitan}
\email{levitanb@physics.mcgill.ca}
\affiliation{Department of Physics, McGill University, 3600 rue University, Montr\'{e}al, Quebec H3A 2T8, Canada}
\author{A.~Metelmann}
\affiliation{Department of Physics, McGill University, 3600 rue University, Montr\'{e}al, Quebec H3A 2T8, Canada}
\affiliation{Department of Electrical Engineering, Princeton University, Princeton, New Jersey 08544, USA}
\author{A.~A.~Clerk}
\affiliation{Department of Physics, McGill University, 3600 rue University, Montr\'{e}al, Quebec H3A 2T8, Canada}

\begin{abstract}
	We consider the physics of an optomechanical cavity subject to coherent two-phonon driving, i.e.\@ degenerate parametric amplification of the mechanical 	mode. We show that in such a system, the cavity mode can effectively ``inherit"  parametric driving from the mechanics, yielding phase-sensitive 		
	amplification and squeezing of optical signals reflected from the cavity. We also demonstrate how such a system can be used to perform single-quadrature 
	detection of a near-resonant narrow-band force applied to the mechanics with extremely low added noise from the optics. The system also exhibits strong differences
	from a conventional degenerate parametric amplifier: in particular, the cavity spectral function can become negative, indicating a negative effective photon temperature.
\end{abstract}

\maketitle

%%%%%%%%%%%%%%%%%%%%%%%%%%%%%%%%%%%%%%%%%%%%%%%%%%%%%%
\section{Introduction}

The field of cavity optomechanics \cite{aspelmeyer_review} has experienced dramatic progress in recent years, spurred onwards both by fundamental interest in 
macroscopic quantum phenomena, as well as the promise of practical applications such as optical amplification (e.g.\@ \cite{massel_nature_microwave_amp, Korppi:2016ampandfreqconversion, Toth:2016dissipationengineering}), optical squeezing (e.g.\@ \cite{Mancini:1994squeezing, Purdy:2013squeezing, kronwald_dissipative_squeezing, Qu:2015squeezing, Kilda:2015squeezing}), and high-sensitivity force detection (e.g.\@ \cite{Clerk:2008bae, Hertzberg:2010bae, Woolley:2013twomodebae, Schreppler:2014sql, Motazedifard:2016forcesensing}).  Almost all experiments are well-described by the linearized theory of optomechanics, in which optical fluctuations (both quantum and classical) are treated as being small in comparison to the classical coherent intracavity amplitude.  This linearized theory has been studied in depth by many authors; one may be forgiven for thinking that there
remains nothing left to say about it.  
%there are no remaining unturned corners in this regime of linearized dynamics.

In this paper, we show that the linearized regime does in fact hold at least a few remaining surprises.  We 
start with a standard setup, wherein an optomechanical cavity (in the good cavity limit) is strongly driven at the red mechanical sideband.  We then add something less standard:  a degenerate ``two-phonon" parametric drive applied to the mechanics.  Such a drive could be realized by e.g.\@ parametrically modulating the spring constant of the mechanical resonator at twice the resonator's natural frequency.  We show here that such a setup provides a unique platform for generating phase-sensitive optical amplification and squeezing; moreover, the resulting physics is {\it not} simply equivalent to having an effective optical degenerate parametric amplifier (DPA).  This ultimately stems from the fact that in our system, amplification and squeezing are obtained by using the optical mode to stabilize the mechanics in a regime of mechanical parametric driving that would otherwise be unstable.

Among the many possible advantages of our system is the fact that the amplification and squeezing can be nearly quantum-limited even when the mechanical environment is far from zero temperature --- while the cavity inherits amplification and squeezing interactions from the mechanics, the mechanical fluctuations are simultaneously cooled by the red-sideband laser drive.  This is in stark contrast to the simplest optomechanical amplifier, realized by a simple blue-sideband cavity drive \cite{massel_nature_microwave_amp}.  Further, the (quadrature-sensitive) parametric amplification of the mechanical response to external forces allows one to directly improve the measurement of such forces, beyond the bound set by the quantum limit on continuous position detection (e.g.\@  \cite{Caves:1980rmp, clerk_RMP}).  Note that though others have previously studied optomechanical systems subject to mechanical parametric driving \cite{Szorkovszky:2011mechsqueezing, Szorkovszky2013, Farace:2012parametricmodulation, Szorkovszky:2014detunedqnd, Lemonde:2015nonlinearityenhancement, Bienert:2015modulatedcooling}, the utility of such an approach in generating optical squeezing and amplification appears to have gone unrecognized.

%%%%%%%%%%%%
\begin{figure}[t]
	\includegraphics[width=0.90\columnwidth]{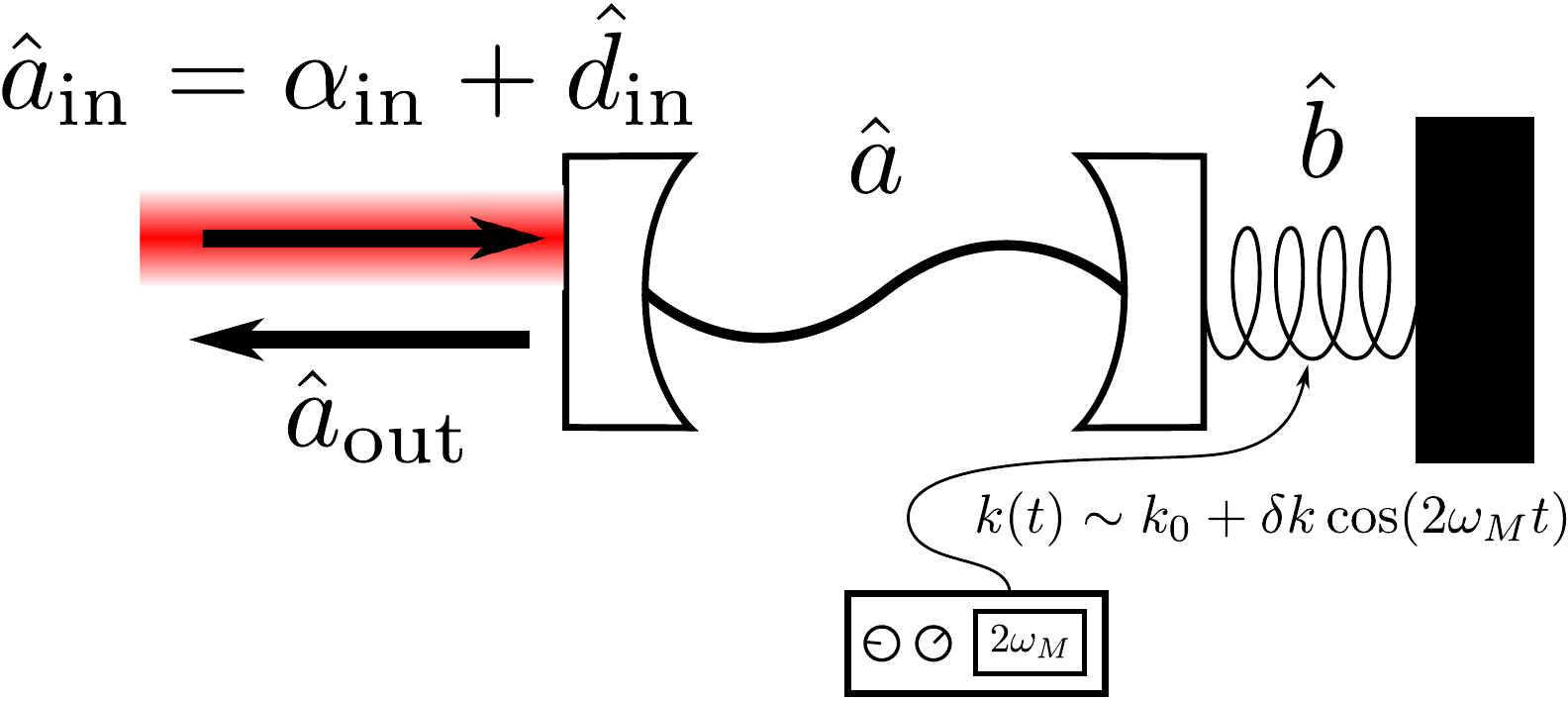}
	\caption{Schematic of the system.  An optomechanical cavity in the good-cavity limit is driven at the red-detuned mechanical sideband, while the
	mechanical resonator is parametrically driven by modulating its spring constant at twice its natural frequency.}
	\label{fig:schematic}
\end{figure}	
%%%%%%%%%%%%

The unusual dynamics in our system also has interesting consequences for an optomechanically induced transparency (OMIT) experiment, where one probes the cavity with a second, weak probe beam \cite{Agarwal:2010omittheory, Weis:2010omit, SafaviNaeini:2011omit}.  Such effects can be tied to an optomechanical modification of the cavity spectral function $A[\omega]$ \cite{Lemonde2013}, which usually plays the role of an effective cavity density of states.  In our system, $A[\omega]$ can become negative, something that is impossible in standard OMIT, or in a standard resonantly-pumped paramp (degenerate or non-degenerate).  We discuss how this implies that the frequency-dependent effective temperature describing the cavity photons becomes negative, indicating a kind of stable population inversion.

%In the case of a nonlinear mechanical resonator with no parametric driving, others \cite{Zhou:2013nonlinearOMIT, Shevchuk:2015nonlinearOMIT} have found that a mechanical Duffing nonlinearity can be mapped onto the cavity OMIT spectrum. 

The remainder of the paper is organized as follows.  In Sec.~\ref{sec:Model}, we introduce the basic model of our system.  Sec.~\ref{sec:ScattAmplification} is devoted to the quantum amplification properties of the system.   We show explicitly that quantum-limited operation is possible even if the mechanical bath temperature corresponds to many thermal quanta.  Sec.~\ref{sec:squeezing} is devoted to the generation of optical squeezing, and provides a detailed comparison against other squeezing protocols, including standard pondermotive squeezing \cite{braginsky_ponderomotive, aspelmeyer_review} and more recent dissipative-squeezing proposals \cite{kronwald_dissipative_squeezing}.  Unlike the pondermotive approach, our system generates squeezing effectively in the good-cavity limit. In Sec.~\ref{sec:ForceSensing}, we discuss how our system can exploit the parametric amplification of one mechanical quadrature to allow the measurement of one quadrature of a mechanical input force with vanishing added measurement noise. Finally, in Sec.~\ref{sec:SpectralFunction}, we discuss OMIT and the unusual behaviour of the cavity spectral function $A[\omega]$, which can become negative.

%%%%%%%%%%%%

%%%%%%%%%%%%%%%%%%%%%%%%%%%%%%%%%%%%%%

%%%%%%%%%%%%%%%%%%%%%%%%%%%%%%%%%%%%%%
\section{Model and linearized theory} \label{sec:Model}

\subsection{Model}
 Our system consists of a driven optomechanical cavity with optical resonance $\omega_c$ and mechanical resonance $\omega_M$, with a parametric drive at $2 \omega_M$ applied to the mechanics. The full Hamiltonian is $\hat{H} = \hat{H}_0 + \hat{H}_{\mathrm{OM}} + \hat{H}_{\mathrm{diss}}$. We begin with the uncoupled cavity mode and mechanical DPA, with coherent Hamiltonian ($\hbar = 1$)
\begin{equation} \label{eq:degenerate_paramp_hamiltonian}
	\hat{H}_0 = \omega_c \hat{a}^{\dagger} \hat{a} + \omega_M \hat{b}^{\dagger} \hat{b}  + \frac{i}{2}\left( \lambda  e^{-2 i \omega_M t} \left( \hat{b}^{\dagger} \right)^2 - h.c. \right).
\end{equation}
$\hat{a}$ ($\hat{b}$) annihilates a photon (phonon), and $\lambda = | \lambda | e^{i \phi_p}$ characterizes the strength and phase of the mechanical parametric driving. The paramp term ($\propto \lambda$) can be realized by e.g.\@ periodic modulation of the spring constant of the mechanical element at $2 \omega_M$ (see, e.g., \cite{Szorkovszky2013}).
%This Hamiltonian can be realized by e.g.\@ periodically modulating the spring constant of the mechanical element at $2 \omega_M$ \textcolor{red}{[CITE SOMETHING]}.
The optical and mechanical modes are coupled via the standard optomechanical interaction
\begin{equation}
	\hat{H}_{\mathrm{OM}} = g \left( \hat{b} + \hat{b}^{\dagger} \right) \hat{a}^{\dagger} \hat{a},
\end{equation}
where $g$ is the single-photon optomechanical coupling.

Dissipation is included via $\hat{H}_{\mathrm{diss}}$, which provides the damping of the cavity and mechanics at rates $\kappa$ and $\gamma$ respectively by independent dissipative baths, and brings in the corresponding noise for each mode. It also provides the driving of the cavity by a coherent source at frequency $\omega_L$.

As we are aiming for the cavity to inherit amplification and squeezing from the mechanics, it is natural to work with a beamsplitter interaction (which can straightforwardly provide state transfer between bosonic modes --- see e.g.\@ \cite{Parkins:1999statetransfer}). Assuming the good-cavity limit $\omega_M \gg \kappa$, such an effective linear interaction can be obtained from the full nonlinear optomechanical interaction in the usual way.  
Choosing the coherent cavity drive to be on the red sideband ($\omega_L = \omega_c - \omega_M$) and working in an interaction picture with respect to the Hamiltonian $\omega_c \hat{a}^{\dagger} \hat{a} + \omega_M \hat{b}^{\dagger} \hat{b}$, we displace away the classical cavity amplitude $\expec{\hat{a}} = \bar{a} = |\bar{a}| \exp{+ i (\omega_M t + \phi_c)}$ by writing $\hat{a} = e^{i \phi_c} ( | \bar{a} | e^{i \omega_M t} + \hat{d} )$, and linearize around the classical solution. This yields the linearized optomechanical interaction
\begin{equation} \label{eq:linearized_interaction}
	\hat{H}_{\mathrm{OM}} = G \left( \hat{d}^{\dagger} \hat{b} + \hat{b}^{\dagger} \hat{d} \right) + \hat{H}_{\mathrm{CR}}.
\end{equation}
$G = g | \bar{a}| $ is the many-photon optomechanical coupling. 
%Note that by writing $\hat{a} = e^{i \phi_c} (| \bar{a} | + \hat{d})$, we work in a gauge where $\hat{X}^{(0)} \equiv (\hat{d} + \hat{d}^{\dagger})/\sqrt{2}$ and $\hat{Y}^{(0)} \equiv (\hat{d} - \hat{d}^{\dagger})/(\sqrt{2} i)$ describe fluctuations in the intracavity amplitude and phase quadratures respectively. 

\begin{figure}[t]
	\centering	
	\includegraphics[width=0.45\textwidth]{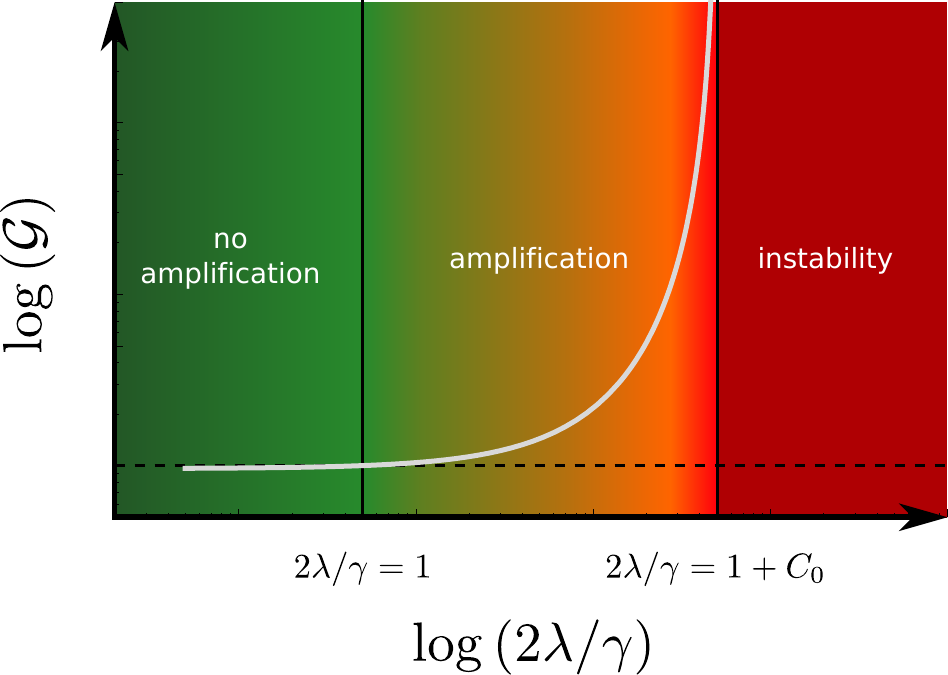}
	\caption{Heuristic behaviour of the optical photon-number gain $\mathcal{G}$ as a function of the mechanical parametric drive amplitude $\lambda$.  
	Optical amplification occurs in the regime where the parametric drive overwhelms the intrinsic mechanical damping, i.e.\@ $|\lambda| > \gamma  / 2$. The system becomes unstable once the parametric drive overwhelms the total mechanical damping, consisting of the intrinsic damping $\gamma$ and the optomechanical damping $C_0 \gamma$. Note that we have assumed weak coupling such that $C_0 \gamma < \kappa$.}
	\label{fig:lambdaregimes}
\end{figure}	

As well as providing means for state transfer, the beamsplitter terms in Eq.~\eqref{eq:linearized_interaction} lead to the well-known cavity-cooling effect \cite{marquardt_cooling,wilsonrae_cooling}, wherein the cavity mode serves to damp and cool the mechanical motion. With our choice of red-sideband drive, the counter-rotating $\hat{H}_{\mathrm{CR}} =  G \left( e^{- 2 i \omega_M t}  \hat{d} \hat{b} + h.c. \right)$ describes off-resonant processes which are strongly suppressed in the good-cavity limit $\kappa \ll \omega_M$. We focus on the good-cavity limit for simplicity, but include $\hat{H}_{\mathrm{CR}}$ in plots unless otherwise noted. For details see Appendix \ref{sec:beyond_RWA_appendix}.

The effective mixing of the mechanical parametric drive with the cavity drive creates a phase reference at the cavity resonance frequency, and determines the phases
of the squeezed and amplified quadratures. We will show that squeezing and amplification are observed in the output light when driving the quadratures
\begin{subequations}		\label{eq:rotated_quadratures}
	\begin{equation}
		\hat{X} \equiv ( e^{-i \phi_p/2} \hat{d} + h.c. ) / \sqrt{2}
	\end{equation}
	and
	\begin{equation}
		\hat{Y} \equiv ( e^{-i \phi_p/2} \hat{d} - h.c. ) / (\sqrt{2} i)
	\end{equation} 
\end{subequations}
respectively.  We stress that these quadratures are defined with respect to the cavity resonance frequency, and not the laser drive frequency.
Note that by varying the paramp phase $\phi_p$ (see text immediately following Eq.~\eqref{eq:degenerate_paramp_hamiltonian}), one can obtain squeezing or amplification of any desired quadrature.  
%Further details on how to obtain the needed relative phase stability is discussed in Appendix \ref{sec:phase_appendix}.

%In the following, we take real $\lambda > 0$ (i.e.\@ $\phi_p = 0$) for convenience; the squeezed quadrature will then be the cavity amplitude quadrature.

In order to achieve degenerate parametric amplification of signals incident on the cavity, one needs a process which creates two photons ($\propto ( \hat{d}^{\dagger} )^2$). 
Heuristically, the Hamiltonians given by Eqs.~\eqref{eq:degenerate_paramp_hamiltonian} and \eqref{eq:linearized_interaction} can provide just such a process: the paramp acts once, creating two phonons, and the beamsplitter interaction in $\hat{H}_{\mathrm{OM}}$ acts twice, converting these phonons into photons. 
%We will show that the mechanics do indeed provide such an effective parametric interaction for the cavity mode at the level of the equations of motion.

\subsection{Heisenberg-Langevin equations}
We treat the dissipation Hamiltonian $\hat{H}_{\mathrm{diss}}$ using the input-output formalism from quantum optics (see e.g.\@ \cite{clerk_RMP}). The resulting Heisenberg-Langevin equations are
\begin{equation} \label{eq:heisenberg_langevin}
	\dot{\hat{\mu}} = i [ \hat{H}_0 + \hat{H}_{\mathrm{OM}}, \hat{\mu} ]
		- \frac{\Gamma_{\mu}}{2} \hat{\mu} - \sqrt{\Gamma_{\mu}} \hat{\mu}_{\mathrm{in}}
\end{equation}
and their Hermitian conjugates, where $\mu = d, b$, $\Gamma_{d} = \kappa$, and $\Gamma_b = \gamma$. Note that we work in the interaction picture, where $\hat{H}_0$ contains only the mechanical parametric driving term (which is time-independent in this frame). 

In Eq.~\eqref{eq:heisenberg_langevin} we have introduced the zero-mean noise operators $\hat{\mu}_{\mathrm{in}}$. Their non-zero correlators are given by $\expec{ \hat{d}_{\mathrm{in}} (t) \hat{d}^{\dagger}_{\mathrm{in}} (t^{\prime})} = \expec{ \hat{d}^{\dagger}_{\mathrm{in}} (t) \hat{d}_{\mathrm{in}} (t^{\prime}) } + \delta (t - t^{\prime}) = \delta (t - t^{\prime}) \left( \nbarTc + 1 \right)$ and analogously for $\hat{b}_{\mathrm{in}}$, with $\nbarTc$ replaced by $\nbarTm$. $\nbarTc$ ($\nbarTm$) is the thermal occupancy of the cavity (mechanical) bath. As shown in Appendix \ref{sec:stability_appendix}, stability requires $| \lambda | < \mathrm{min} \left\lbrace \frac{\gamma}{2} \left( 1 + C_0 \right), \frac{\gamma + \kappa}{2} \right\rbrace$ where we have introduced the cooperativity $C_0 = 4G^2 / (\kappa \gamma)$. We assume that $(G/\kappa)^2 \ll 1$, which means that the relevant stability condition is 
\begin{equation} \label{eq:stability}
	| \lambda |< (\gamma / 2)(1 + C_0) \equiv \lambda_{\mathrm{max}}. 
\end{equation}
Intuitively, the system is stable provided that the parametric driving does not overwhelm the total mechanical damping, which is the sum of the intrinsic mechanical damping $\gamma$ and the optical damping $C_0 \gamma = 4G^2 / \kappa$ --- the optical damping allows for stronger mechanical parametric pumping than would otherwise be possible without reaching instability. We will see that this extended stability regime (i.e.\@ $\gamma/2 < | \lambda | < (\gamma/2) (1 + C_0)$) is precisely the regime where amplification and squeezing occur (see Fig.~\ref{fig:lambdaregimes}).

%Note that as the cavity damping typically dominates the mechanical damping (i.e.\@ $\kappa \gg \gamma$), the optical damping $C_0 \gamma$ may therefore overwhelm the intrinsic damping $\gamma$ even when the coupling is not strong ($4G < \kappa$).

\subsection{Cavity self-energy and effective squeezing interaction}
As a heuristic first look at the dynamics of our system, we can examine the equations for the cavity mode resulting from the algebraic elimination of $\hat{b}$ from the Fourier-transformed Heisenberg-Langevin equations Eq.~\eqref{eq:heisenberg_langevin}. Neglecting noise terms, one obtains
\begin{equation} \label{eq:cavity_EOM}
	-i \omega \hat{d} [\omega] = - \left( \frac{\kappa}{2} + i \Sigma_d [\omega] \right) \hat{d} [\omega] 
		 + \tilde{\lambda} [\omega]  \hat{d}^{\dagger} [\omega] 
\end{equation}
and its Hermitian conjugate, where
\begin{subequations}		\label{eq:EOM_terms}
	\begin{equation} \label{eq:self_energy}
		\Sigma_d [\omega] = \frac{ G^2 (  \omega + i \gamma / 2)}
			{( \omega +i  \gamma / 2)^2 + | \lambda| ^2}
	\end{equation}
is the cavity self-energy resulting from the optomechanical interaction, and
	\begin{equation}	\label{eq:lambda_tilde}
		\tilde{\lambda} [\omega] =  \frac{G^2 \lambda}{(-i \omega + \gamma / 2)^2 - |\lambda|^2}
	\end{equation}
\end{subequations}
plays the role of an induced (non-local in time) parametric interaction.

In the absence of parametric driving (i.e.\@ when $\lambda = 0$), $\tilde{\lambda} [\omega] = 0$, and the optomechanical modification of the cavity is fully encoded in the cavity self-energy $\Sigma_d [\omega]$ given by Eq.~\eqref{eq:self_energy}. It results in a variety of familiar optomechanical effects, including OMIT. Turning on the parametric drive (i.e.\@ $| \lambda | > 0$), Eq.~\eqref{eq:cavity_EOM} and Eq.~\eqref{eq:lambda_tilde} reveal that the mechanics do indeed mediate a parametric-amplifier-like effective squeezing interaction $\tilde{\lambda} [\omega]$ for the cavity mode. Note that this interaction is frequency-dependent, unlike in a true DPA. 

In addition to producing the sought-after paramp-like term, nonzero $ \lambda $ also modifies the cavity self-energy $\Sigma_{d} [\omega]$. On-resonance, the effective squeezing interaction ($\tilde{\lambda}$) becomes larger than the optomechanically-induced cavity damping ($-2 \mathrm{Im} \Sigma_d$) only when $|\lambda| > \gamma/2$. We hence expect amplification only in this extension of the regime of stability, where the optical damping $C_0 \gamma$ is necessary to stabilize the otherwise-unstable mechanics. Also, as mentioned above, $\Sigma_d [\omega]$ is responsible for OMIT --- we will consider the surprising consequences of nonzero $\lambda$ on OMIT physics in Sec.~\ref{sec:SpectralFunction}.

%For another physical perspective, it is instructive write the Hamiltonian in terms of quadratures. We define quadratures $\hat{X} = ( \hat{d} + \hat{d}^{\dagger} ) / \sqrt{2}$ and $\hat{Y} = ( \hat{d} - \hat{d}^{\dagger} ) / (\sqrt{2} i)$, and denote the corresponding mechanical quadratures by $\hat{U}$ and $\hat{V}$ respectively. In terms of these operators, the RWA Hamiltonian is
%\begin{equation} \label{eq:RWA_quadrature_Hamiltonian}
%	\hat{H}_{\mathrm{RWA}} = G \left( \hat{X} \hat{U} + \hat{Y} \hat{V} \right) 
%		+ \frac{\lambda}{2} \left( \hat{U} \hat{V} + \hat{V} \hat{U} \right)
%		+ \hat{H}_{\mathrm{diss}}.
%\end{equation}
%Heuristically, the paramp term tries to amplify $\hat{U}$, and the beamsplitter term transfers the excess energy from $\hat{U}$ (which would to be unstable when $\lambda > \gamma / 2$ if not for the optomechanical interaction) into $\hat{Y}$, yielding amplification of that quadrature.

%%%%%%%%%%%%%%%%%%%%%%%%%%%%%%%
\section{Scattering and amplification}\label{sec:ScattAmplification}
%The preceding discussion has been concerned with the intracavity field fluctuations $\hat{d}$. 
To evaluate the usefulness of our system as a squeezer/amplifier, we must turn our attention to the \textit{output} light produced by scattering a weak probe off of the cavity. 
It is convenient to work in a basis of quadrature operators: the cavity quadratures $\hat{X}$ and $\hat{Y}$ are defined according to Eqs.~\eqref{eq:rotated_quadratures}, and the analogous mechanical quadratures are denoted by $\hat{U}$ and $\hat{V}$. These four quadratures are collected into the vector $\hat{\matr{Q}} = (\hat{X}, \hat{Y}, \hat{U}, \hat{V})^T$. The scattering matrix $\matr{s}[\omega]$ then links the inputs and outputs according to $\hat{\matr{Q}}_{\mathrm{out}}[\omega] = \matr{s}[\omega] \hat{\matr{Q}}_{\mathrm{in}}[\omega]$. $\matr{s} [\omega]$ can be straightforwardly calculated using input-output theory (see Appendix \ref{sec:scattering_appendix}). On-resonance, it is given by
%\AC{Something wrong here when you take the $C_0 \rightarrow 0$ limit...}
%\begin{multline} \label{eq:resonant_scattering}
%	\matr{s}[0] = \\
%	\begin{pmatrix}
%		\frac{C_+ - 1}{C_+ + 1} & 0 & 0 & -\frac{2 C_+}{\sqrt{C_0}} \frac{1}{C_+ +1} \\
%		0 & \frac{C_- - 1}{C_- + 1}  &  \frac{2 C_-}{\sqrt{C_0}} \frac{1}{C_- +1} & 0 \\
%		0 & -\frac{2 C_-}{\sqrt{C_0}} \frac{1}{C_- +1}  & \frac{C_- + 1 - \frac{2 C_-}{C_0}}{C_- + 1} & 0 \\
%		\frac{2 C_+}{\sqrt{C_0}} \frac{1}{C_+ + 1} & 0 & 0 & \frac{C_+ + 1 - \frac{2 C_+}{C_0}}{C_+ + 1}
%	\end{pmatrix}
%\end{multline}
\begin{multline} \label{eq:resonant_scattering}
	\matr{s}[0] = \\
	\begin{pmatrix}
		\frac{C_0 - 1 - \mathcal{R}_Y}{1 + \mathcal{R}_Y(1 + C_0)} & 0 & 0 & \frac{-\sqrt{C_0} (1 + \mathcal{R}_Y)}{1 + \mathcal{R}_Y (1 + C_0)} \\
		0 & \mathcal{R}_Y & \frac{1 + \mathcal{R}_Y}{\sqrt{C_0}} & 0 \\
		0 & - \frac{1 +\mathcal{R}_Y}{\sqrt{C_0}} & 1 - \frac{1 + \mathcal{R}_Y}{C_0} & 0 \\
		\frac{\sqrt{C_0} (1 +\mathcal{R}_Y)}{1 + \mathcal{R}_Y(1 + C_0)} & 0 & 0 & \frac{C_0 \mathcal{R}_Y}{1 + \mathcal{R}_Y (1+ C_0)}
	\end{pmatrix}.
\end{multline}
This result is parametrized by the previously introduced cooperativity $C_0$, and the resonant $\hat{Y}$-quadrature amplitude reflection coefficient $\mathcal{R}_Y$, i.e.\@ the $Y$-$Y$ element of $\matr{s} [0]$:
\begin{equation} \label{eq:Y_amplitude_gain}
	\mathcal{R}_Y
		\equiv \frac{ C_0 - (1 - 2 | \lambda | / \gamma) }
		{C_0 + (1 - 2 | \lambda | / \gamma)}.
\end{equation}
The photon-number gain for optical signals in the $\hat{Y}_{\mathrm{in}}$-quadrature is then $\gain = \abs{\mathcal{R}_Y}^2$.
Note that precisely as expected based on our earlier analysis of the intracavity dyamics, above-unity gain occurs only when $| \lambda | > \gamma / 2$ --- the (unstable) regime of parametric oscillation for an uncoupled mechanical resonator.  Combined with the stability condition Eq.~\eqref{eq:stability}, this means that stable amplification of the electromagnetic $Y$ quadrature occurs in the optically-stabilized regime $\gamma / 2 < | \lambda | < (\gamma / 2) (1 + C_0)$, as illustrated in Fig.\@ \ref{fig:lambdaregimes}. 

\begin{figure}
	\includegraphics[width=0.45\textwidth]{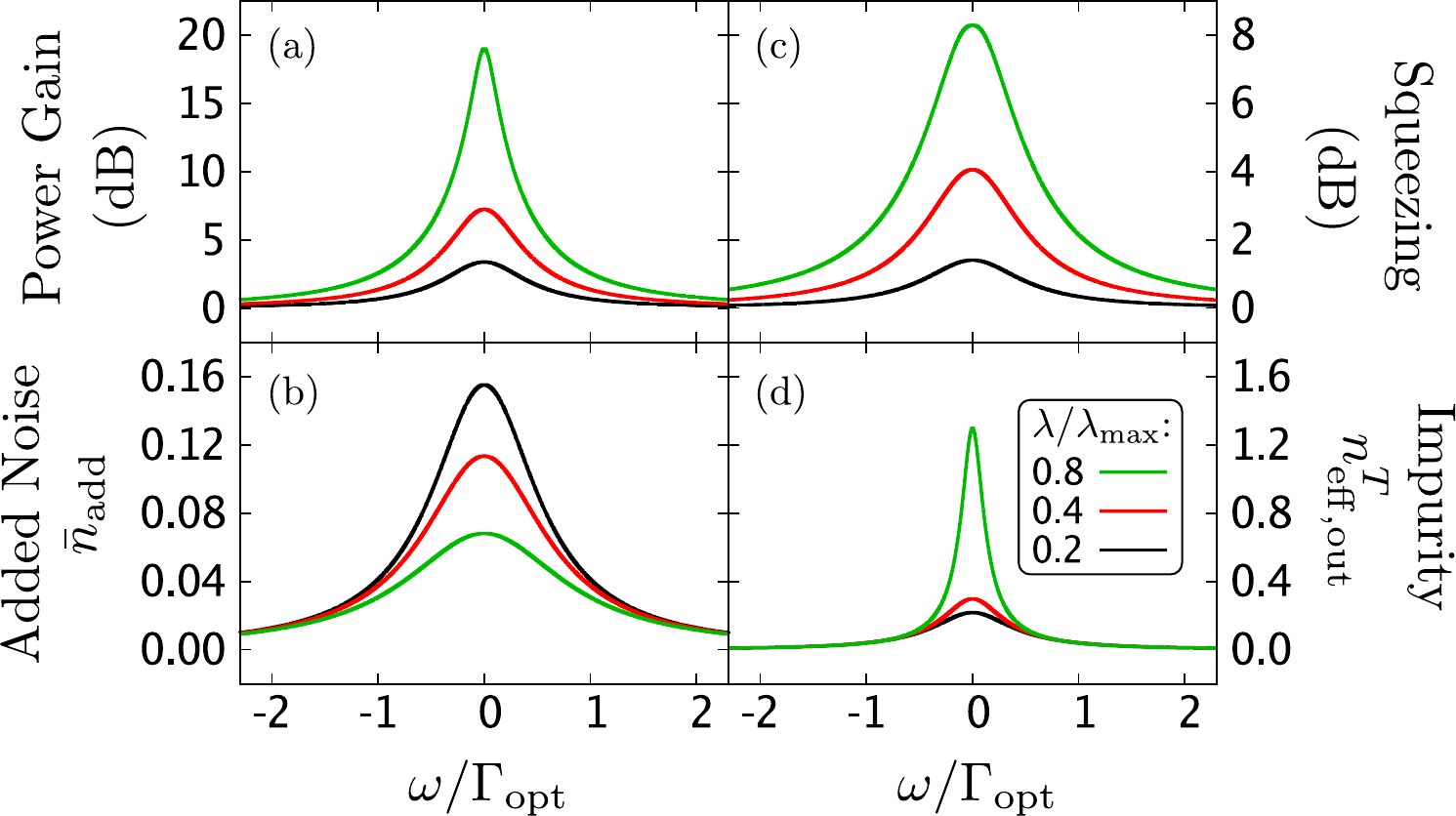}
	\caption{Optical amplification and squeezing properties. (a) Power gain and (b) symmetrized added noise spectral density (referred back to the input) for the optical
	$\hat{Y}_{\mathrm{out}}[\omega]$ output quadrature. (c) Squeezing of the $\hat{X}_{\mathrm{out}}[\omega]$ quadrature below zero-point. (d) Impurity of the cavity output. 
	For all plots, $ \gamma / \kappa = 10^{-4}$, $G / \kappa = 5 \times 10^{-2}$ (such that $C_0 = 100$), and $\omega_M / \kappa = 20$; we keep first-sideband non-RWA corrections. The optical bath is at zero temperature, and the mechanical bath temperature corresponds to $\bar{n}^T_m = 5$. Increasing the parametric drive $\lambda$ towards $\lambdamax = (\gamma/2)(1+C_0)$ yields increased gain, increased squeezing and reduced added noise (referred back to the input), while sacrificing purity. $\Gamma_{\mathrm{opt}} \equiv C_0 \gamma$ is the optical damping rate.}
	\label{fig:gain_and_noise}
\end{figure}

In an ordinary DPA, the added noise in the amplified quadrature disappears in the large-gain limit. Surprisingly, despite the involvement of a second mode (the mechanics), our system can approach this ideal behaviour. From the scattering Eq.~\eqref{eq:resonant_scattering} expressed in terms of the amplitude gain $\sqrt{\gain}$ and the cooperativity $C_0$ (assuming $C_0 > 1$ so that $\sqrt{\gain} = + \mathcal{R}_Y$), the total noise power in $\hat{Y}_{\mathrm{out}} [0]$ referred back to the input is given by
%\begin{equation} \label{eq:resonant_added_noise}
%	\frac{|s_{YU}[0]|^2}{|s_{YY}[0]|^2} \left( \nbarTm + 1/2 \right)
%	= \frac{1}{C_0} \left( 1 + \frac{1}{\sqrt{\gain}} \right)^2  
%		\left( \nbarTm + 1/2 \right),
%\end{equation}
\begin{align} \label{eq:resonant_added_noise}
	\frac{\bar{S}_{YY}^{\mathrm{out}}[0]}{\gain}
	& = \nbarTc + \frac{1}{2} + \frac{1}{C_0} \left( 1 + \frac{1}{\sqrt{\gain}} \right)^2  
		\left( \nbarTm + 1/2 \right) \\
	& \equiv \nbarTc + \frac{1}{2} + \nAddAmp [0],
\end{align}
where for operators $\hat{A}_{\mathrm{out}}$ and $\hat{B}_{\mathrm{out}}$,
\begin{equation}	\label{eq:define_power_spectrum}
	\bar{S}^{\mathrm{out}}_{AB} [\omega] = \frac{1}{2} \int \mathrm{d} t \, e^{i \omega t} 
		\expec{ \lbrace \hat{A}_{\mathrm{out}} (t), \hat{B}_{\mathrm{out}} (0) \rbrace }
\end{equation}
is the symmetrized (i.e.\@ classical) correlator.
$\nAddAmp$ is the standard amplifier added noise (referred back to the input), expressed as a number of quanta. Notice that this noise, originating from the mechanical bath, is cavity-cooled, and disappears as $C_0 \rightarrow \infty$. Indeed, in the limit where $\gain$ is held fixed while $C_0 \rightarrow \infty$, one has
\begin{equation} \label{eq:limit_scattering_matrix}
	\matr{s}[0] \rightarrow
	\begin{pmatrix}
		\frac{1}{\sqrt{\gain}} & 0 & 0 & 0 \\
		0 & \sqrt{\gain} & 0 & 0 \\
		0 & 0 & 1 & 0 \\
		0 & 0 & 0 & 1
	\end{pmatrix}.
\end{equation}
This is precisely the scattering behaviour of a quantum-limited phase-sensitive amplifier \cite{caves_limits} which is entirely decoupled from the mechanics. The added noise for large but realistic cooperativities (e.g.\@ $C_0 \sim 100$) and non-zero mechanical bath temperature is shown in Fig.\@ \ref{fig:gain_and_noise} (b). The suppression of mechanical noise in the amplifier output, which can also be provided by the dissipative optomechanical amplification scheme \cite{Metelmann:2014dissipativeamp}, stands in stark contrast to the behaviour of the simplest optomechanical amplifier, the non-degenerate paramp (NDPA) realized by driving an optomechanical cavity on its blue sideband  \cite{massel_nature_microwave_amp}. In such an NDPA the mechanical noise is not cooled, and can represent a significant source of added noise for the amplifier.

%\section{Bandwidth}
As is the case for other flavours of parametric amplifier, our scheme is subject to a gain-bandwidth limitation. This limitation can be straightforwardly obtained from the frequency-dependent scattering matrix (see Appendix \ref{sec:scattering_appendix}). For large gain, large $C_0$ and large $\kappa / G$, the amplification bandwidth (i.e.\@ the FWHM of $|s_{YY} [\omega]|^2$) is well-approximated by
\begin{equation} \label{eq:approx_bandwidth}
	B \approx (1/ \sqrt{\gain}) \left( 8G^2 / \kappa \right).
\end{equation}
%\begin{equation} \label{eq:approx_bandwidth}
%	B \approx \frac{1}{\sqrt{\gain}} \frac{8G^2}{\kappa}.
%\end{equation}
The gain-bandwidth product for our system is thus controlled by the optical damping $C_0 \gamma  = 4G^2 / \kappa$.  This again compares favourably against the optomechanical amplifier of Ref.~\onlinecite{massel_nature_microwave_amp}, where the gain-bandwidth product is limited by the much smaller mechanical damping rate $\gamma$.
Note that recent experiments \cite{Korppi:2016ampandfreqconversion, Toth:2016dissipationengineering} have investigated multi-mode approaches to optomechanical amplification which lead to improved amplifier bandwidth \cite{Metelmann:2014dissipativeamp, Nunnenkamp:2014reverseddissipation}. %Unlike a conventional DPA, amplification is not achieved over the full cavity bandwidth $\kappa$ because the mechanically-mediated effective parametric interaction \eqref{eq:lambda_tilde} acts over a smaller bandwidth. 

%Once again, the ability to cavity-cool the mechanical mode while simultaneously using that same mode to mediate amplification for the cavity mode yields a useful improvement in performance, in this case increasing the effective bandwidth-gain product by a potentially substantial factor of $C_0$.

\section{Squeezing}	\label{sec:squeezing}

\subsection{Squeezing generation}

\begin{figure}
	\includegraphics[width=0.45\textwidth]{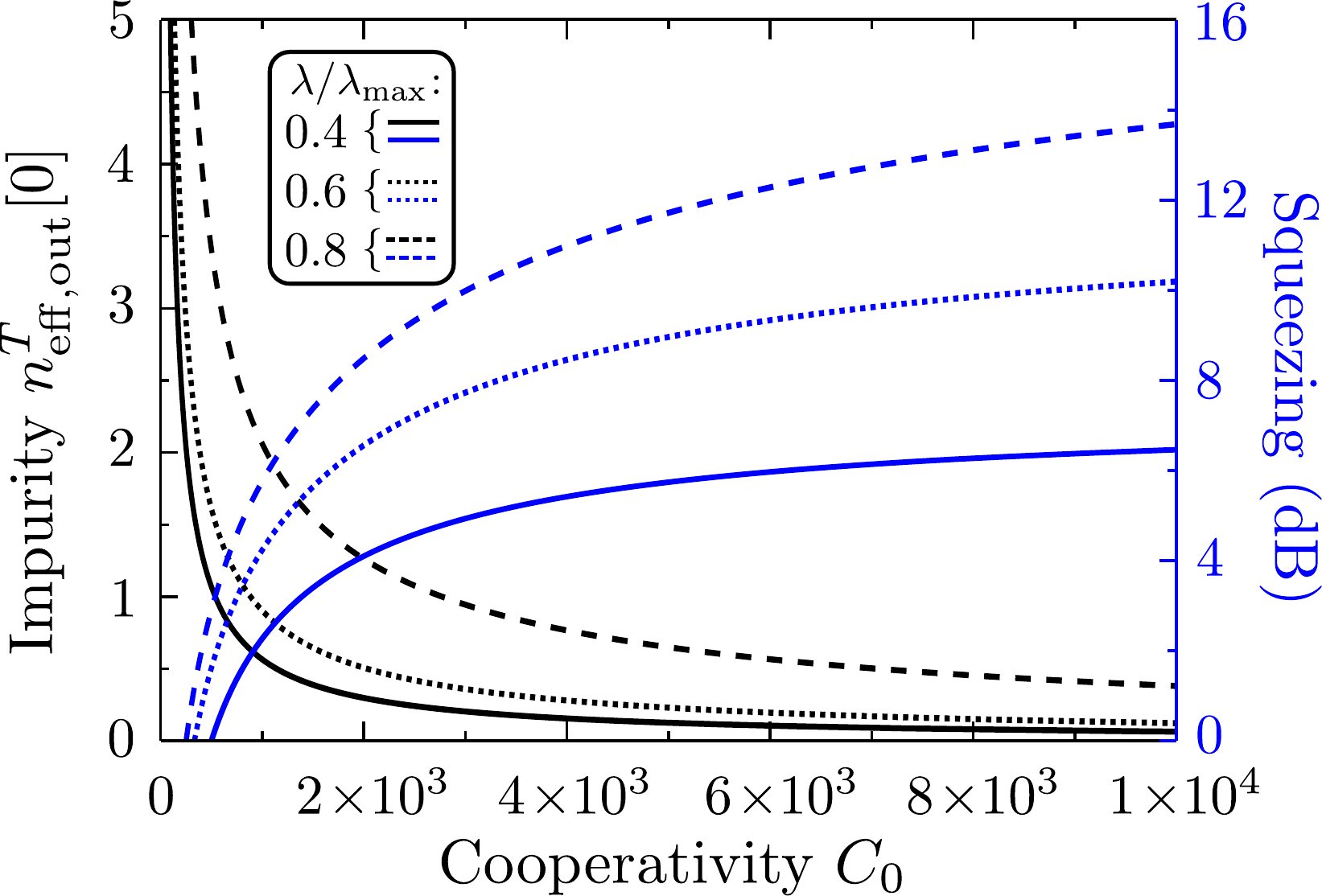}
	\caption{Squeezing and purity of the output light on-resonance. Blue curves show squeezing of $\bar{S}^{\mathrm{out}}_{XX} [0]$ expressed in dB below zero-point. Black curves show the associated impurity of the output light. Here we take $ \gamma / \kappa = 10^{-5}$, $\omega_M / \kappa = 20$, $\bar{n}^T_c = 0$, and $\bar{n}^T_m = 100$.}
	\label{fig:squeezing_vs_coop}
\end{figure}

Noiseless phase-sensitive amplification of one quadrature goes hand-in-hand with squeezing of its complementary quadrature. In keeping with this, our scheme is capable of producing significant squeezing of the cavity output field. For large cooperativities and a zero-temperature cavity input, the squeezing of the on-resonance cavity output quadrature $\hat{X}_{\mathrm{out}} [0]$ below zero-point  is given by
\begin{align} 
	e^{-2r} & \equiv \frac{ \bar{S}_{XX}^{\mathrm{out}}[0]  }{ 1 /2 } 	\label{eq:resonant_squeezing_large_C_a}
	\\ & \approx 1 - \frac{4 ( |\lambda| / \lambdamax)  (1-1/C_0) }{\left( 1 + | \lambda | / \lambdamax \right)^2}
		+ \frac{8 \nbarTm}{C_0 \left( 1 + | \lambda | / \lambdamax \right)^2} 	\label{eq:resonant_squeezing_large_C_b}
	\\ & \xrightarrow{| \lambda | \rightarrow \lambda_{\mathrm{max}}} \frac{ 2 \nbarTm + 1}{C_0} 	\label{eq:resonant_squeezing_large_C_c},
\end{align}
where $\lambda_{\rm max}$ is defined in Eq.~(\ref{eq:stability}).

Note that the maximum degree of squeezing is set by the cavity-cooled mechanical temperature; significant squeezing below zero-point of the cavity output field therefore requires the same magnitude of cooperativity as is needed to approach the mechanical ground state via optomechanical sideband cooling.  Typical squeezing versus cooperativity
curves are shown in Fig.\@ \ref{fig:squeezing_vs_coop}.
%A recent experiment \cite{Peterson:2016cooling} demonstrated cooling to approximately $0.2$ thermal phonons, beginning from an equilibrium value of $\nbarTm \sim 10^3$. This corresponds to an approximate cooperativity of $C_0 \sim 5 \times 10^3$. With our method, these parameters could enable squeezing to a maximum of $e^{-2r} \approx 0.4 < 1$.

In addition to the amount of squeezing, the purity of that squeezing is an important figure of merit. The impurity of the cavity output may be quantified by an effective thermal occupancy $\nTeffout$, defined via
\begin{equation} \label{eq:define_effective_thermal_occupancy}
	(\nTeffout [\omega] + 1/2)^2 = \bar{S}^{\mathrm{out}}_{XX} [\omega] 
		\bar{S}^{\mathrm{out}}_{YY} [\omega] -  
		\bar{S}^{\mathrm{out}}_{XY} [\omega] \bar{S}^{\mathrm{out}}_{YX} [\omega].
\end{equation}
$\nTeffout$ thus defined will be zero for any pure state of the output light, and equal to the actual thermal occupancy for a thermal state (see e.g.\@ \cite{braunstein_RMP}). For our system in the RWA, the cross-correlators between $\hat{X}_{\mathrm{out}}$ and $\hat{Y}_{\mathrm{out}}$ vanish, leaving only the diagonal term in Eq.~\eqref{eq:define_effective_thermal_occupancy}. In the large-cooperativity limit $C_0 \gg 1$ and on-resonance, one has (to order $1 / C_0$)
\begin{multline}	\label{eq:impurity}
	(\nTeffout [0] + 1/2)^2
	\\ \approx \frac{1}{4} + \frac{4}{C_0} \frac{ \abs{ \frac{\lambda}{\lambdamax} }^2 + 
		\nbarTm \left( 1 + \abs{ \frac{\lambda}{\lambdamax} }^2 \right)}
		{ \left( 1 - \abs{ \frac{\lambda}{\lambdamax} }^2 \right)^2}.
\end{multline}
%\begin{multline} VERSION INCLUDING (n^T_m / C)^2
%	(\nTeffout [0] + 1/2)^2
%	\\ \approx \frac{1}{4} + \frac{4}{C_0} \frac{ \left( \frac{\lambda}{\lambdamax} \right)^2 + 
%		\nbarTm \left( 1 + \left( \frac{\lambda}{\lambdamax} \right)^2 + \frac{4 \nbarTm}{C_0} \right)}
%		{ \left( 1 - \left( \frac{\lambda}{\lambdamax} \right)^2 \right)^2}
%\end{multline}

As maximal squeezing occurs when $|\lambda| \rightarrow \lambdamax$, there is thus a tradeoff between the degree of squeezing achieved and the purity of that squeezing. In a serendipitous accident of terminology, the degree of achievable compromise is controlled by the cooperativity: Larger cooperativities allow greater purity for a given amount of squeezing, as can be seen in Fig.\@ \ref{fig:squeezing_vs_coop}. As the system approaches instability, fluctuations in the mechanical $\hat{U}$ quadrature are amplified by the parametric driving but cooled by the red-sideband interaction with the cavity mode. As these fluctuations are then transferred into the cavity $\hat{Y}$ quadrature by way of the optomechanical interaction, the degree of impurity of the cavity output field reflects the competition between these heating and cooling effects. 
%Note that this parametric amplification of mechanical fluctuations does not impact the degree of squeezing of the cavity output quadrature $\hat{X}_{\mathrm{out}}$ (see \eqref{eq:resonant_squeezing_large_C_c}) because the paramp only amplifies the fluctuations in the mechanical $\hat{U}$ quadrature, which couples only to the cavity $\hat{Y}$ quadrature.
% However, since $\nTeffout$ knows about both quadratures of the output field, these amplified fluctuations \textit{do} have an effect on the cavity output purity as indicated by \eqref{eq:impurity}.

\subsection{Comparison against other squeezing protocols}

\captionsetup[subfigure]{justification=raggedright}
\begin{figure*}
	\centering
	\begin{subfigure}[t]{0.49\linewidth}
		\caption{}
		\includegraphics[width=\linewidth]{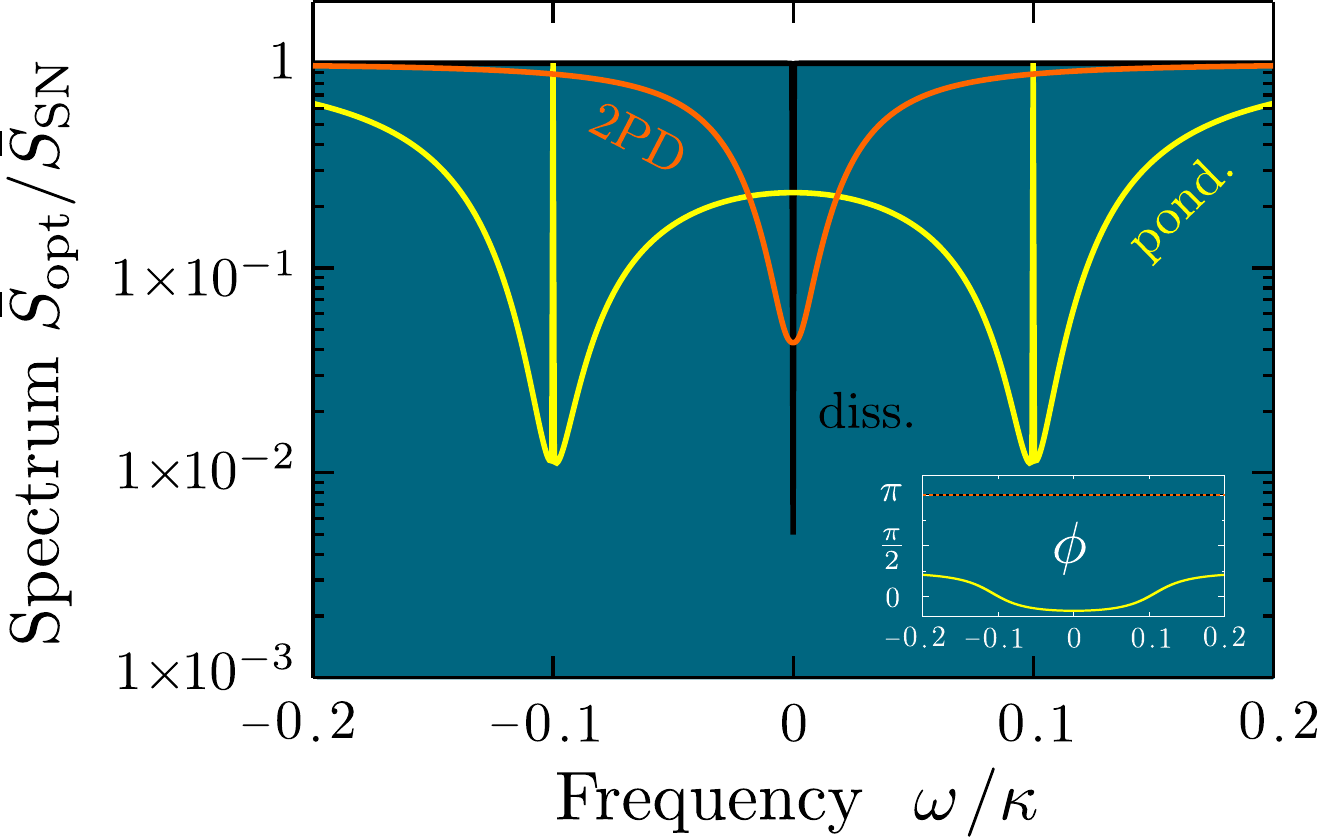}
		\label{fig:squeezing_comparison}
	\end{subfigure}
	\begin{subfigure}[t]{0.49\linewidth}
		\caption{}
		\includegraphics[width=\linewidth]{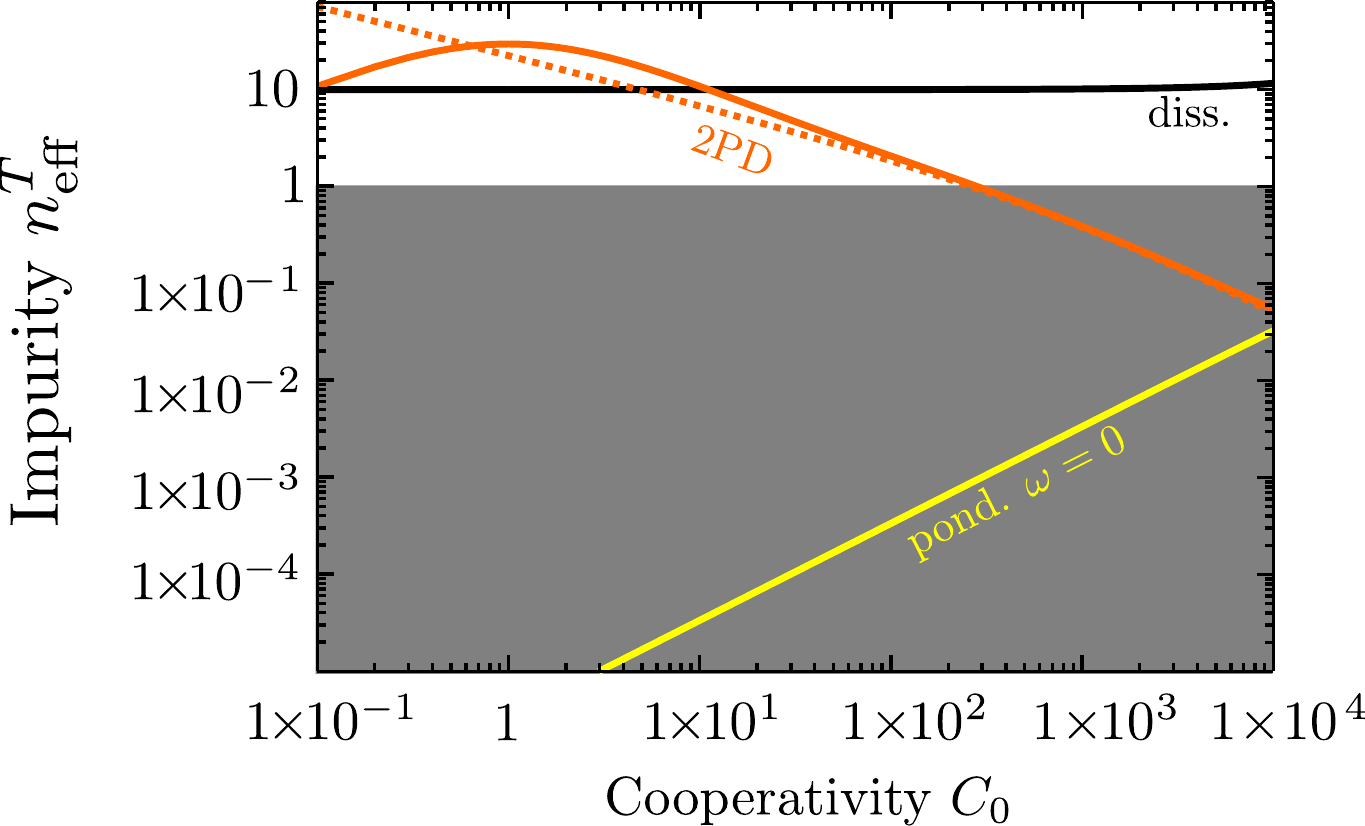}
		\label{fig:purity_comparison}
	\end{subfigure}
	\caption{Comparison between different optomechanical squeezing protocols: ``2PD" denotes the two-phonon drive setup of this paper, ``diss." corresponds to the dissipative scheme \cite{kronwald_dissipative_squeezing} and ``pond." corresponds to standard ponderomotive squeezing with a resonant drive (e.g.\@  \cite{aspelmeyer_review}).  (a) The noise spectrum for the maximally-squeezed quadrature angle $\phi$ in units of zero-point noise; dark blue indicates squeezing below the shot noise level.   Note that $\phi$ varies as a function of frequency in the ponderomotive case only. $C_0 =  1000$ for all cases. While the dissipative scheme produces a greater degree of squeezing than does our approach, our approach enjoys a squeezing bandwidth improvement of a factor $\sim C_0$ relative to the dissipative scheme. (b) Comparing the output state impurity produced by different squeezing schemes in terms of effective thermal quanta (see Eq.~\eqref{eq:define_effective_thermal_occupancy}). Note that the impurity produced by the ponderomotive approach near the mechanical sidebands is too small to display on this scale. The dashed orange line is based on the approximate result in Eq.~\eqref{eq:impurity}.
	In both plots, $\gamma/\kappa = 10^{-5}$, $\nbarTm = 10$, and $\nbarTc = 0$ for all schemes. Our method uses $\lambda / \lambda_{\mathrm{max}} = 0.8$. We take $\omega_M / \kappa = 20$ for our method and the dissipative scheme, and take $\omega_M / \kappa = 0.1$ for the ponderomotive scheme.}
\end{figure*}

%Taking the various requirements together, 
Our scheme produces squeezed output light most efficiently in the good-cavity limit ($\kappa \ll \omega_M$) with weak coupling ($G < \kappa$) and large cooperativity ($C_0 \gg 1$). In contrast, the standard ponderomotive squeezing mechanism \cite{braginsky_ponderomotive, aspelmeyer_review} is efficient only in the \textit{bad}-cavity limit $\kappa \gg \omega_M$. Although the good-cavity limit is desirable for the realization of several optomechanical processes, if one is willing to work in the bad-cavity limit, then the squeezing achieved by ponderomotive squeezing can be significantly more pure than the squeezing achievable using our scheme in the good-cavity limit with a similar $C_0$ (see Fig.\@ \ref{fig:purity_comparison}).  Another significant difference is that in the ponderomotive case, the squeezing angle is dependent on frequency, while in our system it is always the $X$ quadrature of the cavity output which is squeezed (see inset in Fig.\@ \ref{fig:squeezing_comparison}). Recall that the angle defining the $\hat{X}$-quadrature is controlled by the paramp phase $\phi_p$ (see Eq.~\eqref{eq:rotated_quadratures}).
%Recall that the angle between $\hat{X}$ and the cavity amplitude quadrature $\hat{X}^{(0)}$ is controlled by the paramp phase $\phi_p$ (see Eq.~\eqref{eq:rotated_quadratures}).

The dissipative optomechanical squeezing scheme \cite{kronwald_dissipative_squeezing} also yields a frequency-independent squeezing angle, but suffers from a very narrow bandwidth, set by the bare mechanical damping $\gamma$. In contrast, when $\sqrt{\gain}$, $C_0$ and $\kappa / G$ are all large, our scheme yields a squeezing bandwidth controlled by the optical damping $4 G^2  / \kappa$, thus providing a bandwidth improvement by a potentially large factor $C_0$ relative to the dissipative scheme --- see Fig.\@ \ref{fig:squeezing_comparison}. While the dissipative scheme is capable of producing a greater degree of squeezing, the impurity of the cavity output in that case is set by the temperature of the mechanical bath as opposed to the (much lower) cavity-cooled temperature achievable in our scheme provided that one does not drive the system too close to the parametric instability. A detailed comparison between the ponderomotive and dissipative squeezing schemes can be found in \cite{kronwald_dissipative_squeezing}.

%%%%%%%%%%%%%%%%%%%%%%%%%%%%%%%%%%%%%%
\section{Single-quadrature force sensing} \label{sec:ForceSensing}

%\AC{Need to re-write, as a force detector that attenuates the main signal isn't interesting....}

So far we have considered the output light produced by the cavity in response to weak optical inputs, with the mechanics driven only by noise. However, through the optomechanical interaction, mechanical input signals (i.e.\@ forces) are also imprinted onto the cavity output. We have already demonstrated how the second row of the scattering matrix (Eq.~\eqref{eq:resonant_scattering} for the resonant case and Eq.~\eqref{eq:freq_dependent_scattering} for the non-resonant case) leads to quadrature-sensitive amplification of the optical output field. This row, in particular the off-diagonal $Y$-$U$ element, describes the transduction of one quadrature of mechanical input signals, i.e.\@ forces, to the ``amplified" \footnote{For efficient force-sensing, we will choose parameters such that signals in the $\hat{Y}_{\mathrm{in}}$ quadratures are in fact \textit{attenuated}, while the mechanical response to the input force quadrature $\hat{U}_{\mathrm{in}}$ is amplified --- 
we refer here to $\hat{Y}$ as the amplified quadrature only to make clear that this is the same quadrature whose amplification was discussed in Sec.\@ \ref{sec:ScattAmplification}.}  optical output quadrature:  hence, monitoring $\hat{Y}_{\mathrm{out}}$ provides a measurement of $\hat{U}_{\mathrm{in}}$. We will now show that this process allows for single-quadrature force detection with arbitrarily small
added noise, thus allowing one to surpass the usual quantum limit on force detection.

The (classical) mechanical input force is described in the lab frame by a Hamiltonian
\begin{equation}
	\hat{H}_F = F (t) \hat{x}_m = \sqrt{\frac{\hbar}{2 m \omega_M}} F (t) (\hat{b} + \hat{b}^{\dagger}),
\end{equation}
where $F(t)$ is the mechanical force to be detected and $\hat{x}_m$ is the (dimensionful) position of the mechanical oscillator (mass $m$). In terms of the classical force $F$, the mechanical quadrature Fourier component $\hat{U}_{\mathrm{in}} [\omega]$ is given by 
\begin{multline}	\label{eq:detected_force_quadrature}
	\hat{U}_{\mathrm{in}} [\omega] = \hat{U}_{\xi} [\omega] + \frac{1}{2 i}  \sqrt{\frac{\hbar}{2 m \omega_M}} \left( e^{i \phi_p /2} F [\omega + \omega_M] \right. \\
		\left. - e^{- i \phi_p /2} F [\omega - \omega_M] \right),
\end{multline}
where $\hat{U}_{\xi}$ is the zero-mean $U$-quadrature of the mechanical input noise satisfying $\langle \hat{U}_{\xi} (t) \hat{U}_{\xi} (t^{\prime}) \rangle = ( \nbarTm + 1/2) \delta (t - t^{\prime})$. Recall that the mechanical quadratures are defined in a rotating frame, and involve the phase $\phi_p/2$ in their definition.

\begin{figure*}[t]
	\includegraphics[width=\textwidth]{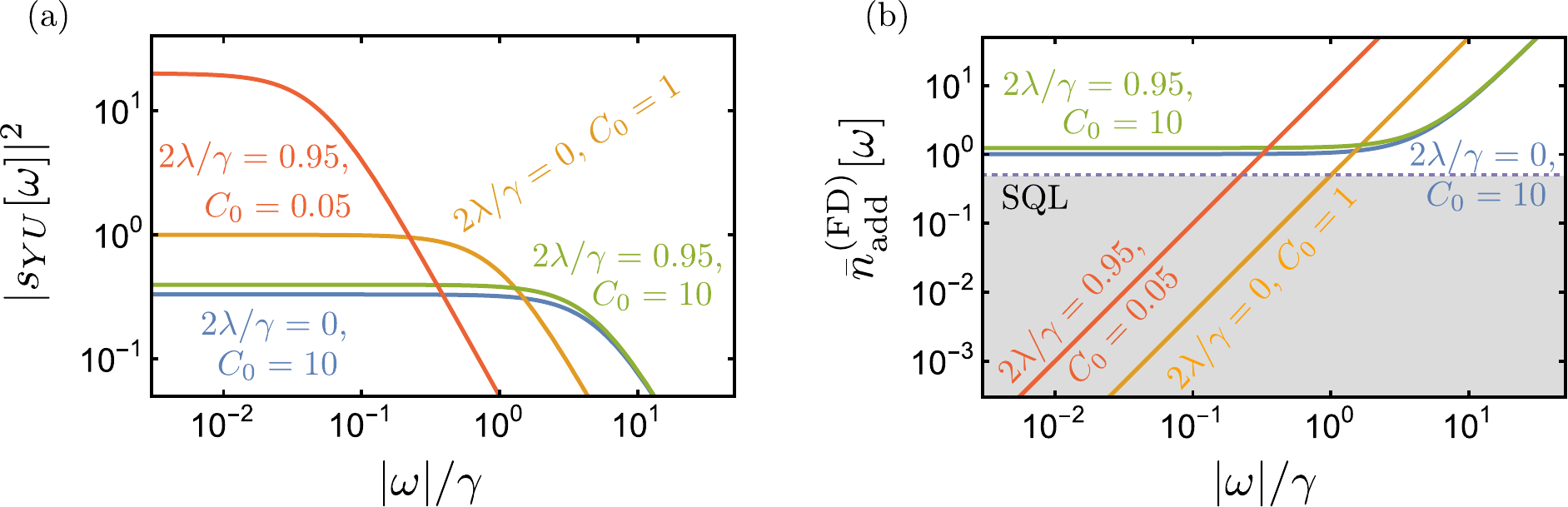}
	\caption{(a) Modulus-square of the scattering coefficient $s_{YU} [\omega]$, describing transduction of the mechanical force quadrature $\hat{U}_{\mathrm{in}} [\omega]$ (see Eq.~\eqref{eq:detected_force_quadrature}) to the optical output quadrature $\hat{Y}_{\mathrm{out}} [\omega]$. (b) Optically-added noise for a measurement of $\hat{U}_{\mathrm{in}} [\omega]$, expressed as an equivalent number of mechanical bath noise quanta accompanying the force to be detected. By satisfying the impedance matching condition $C_0 + 2 \lambda / \gamma = 1$, the optically-added noise can be made to vanish on-resonance while simultaneously providing $\abs{s_{YU} [\omega]} > 1$. This plot takes $\gamma / \kappa = 10^{-5}$, and assumes the RWA.}
	\label{fig:force_sensing}
\end{figure*}

To measure this force quadrature, one must detect the optical output quadrature $\hat{Y}_{\mathrm{out}} (t)$.
% In terms of lab-frame quantities (subscript $_\mathrm{lab}$), it is given by
%\begin{equation}
%	\hat{X}_{\mathrm{out}} (t) = \cos (\omega_c t) \hat{X}_{\mathrm{lab, out}} (t) + \sin (\omega_c t) \hat{Y}_{\mathrm{lab, out}} (t).
%\end{equation}
An important figure of merit in such a measurement is the total added noise of the measurement, which here consists of the contribution of the input optical vacuum noise  
in $\hat{Y}_{\mathrm{out}} (t)$.
%accompanying the transduced mechanical signal. 
This added noise can be viewed as an effective increase in the force fluctuations originating from the mechanical bath. To that end, it is convenient to quantify it as an equivalent number of bath noise quanta $\nAddFD [\omega]$:
\begin{multline}
	\bar{S}_{Y}^{\mathrm{out}} [\omega] = |s_{YY} [\omega]|^2 \left( \nbarTc + \frac{1}{2} \right) + |s_{YU} [\omega] |^2 \left( \nbarTm + \frac{1}{2} \right) \\
		\equiv | s_{YU} [\omega] |^2 \left( \nbarTm + \frac{1}{2} + \nAddFD [\omega] \right).
\end{multline}
%i.e.\@
%\begin{equation}
%	\bar{n}_{\mathrm{imp}} [\omega] = \abs{\frac{s_{XX} [\omega]}{s_{XV} [\omega]}}^2 \left( \nbarTc + \frac{1}{2} \right).
%\end{equation}

%\begin{figure}
%	\includegraphics[width=0.48\textwidth]{force_transduction_coefficient}
%	\caption{Modulus-square of the scattering coefficient $s_{XV} [\omega]$, describing transduction of the mechanical force quadrature $V_{\mathrm{in}} [\omega]$ (see \eqref{eq:detected_force_quadrature}), to the optical output quadrature $\hat{X}_{\mathrm{in}} [\omega]$. The parametric gain is taken to be $\gain \equiv |s_{YY} [0]|^2 = 100$.  This plot takes $\kappa / \gamma = 10^5$, and assumes the RWA.}
%	\label{fig:force_transduction_coefficient}
%\end{figure}

From the on-resonance scattering matrix in Eq.~\eqref{eq:resonant_scattering} and the expression Eq.~\eqref{eq:Y_amplitude_gain}, one finds something remarkable:  
when $C_0 \gamma = \gamma - 2 \abs{\lambda}$ with $\lambda \ne 0$  (which can only happen when $C_0$ and $2 \lambda / \gamma$ are both less than $1$), the on-resonance optically-added noise {\it vanishes exactly}, while, at the same time, the mechanical parametric driving provides an amplified response to the mechanical input force. This vanishing of the optically-added noise can be thought of as resulting from an impedance matching condition for the $\hat{U}$-quadrature:  one has balanced the 
paramp-modified intrinsic mechanical damping of this quadrature, $\gamma - 2 \abs{\lambda}$, against the (phase-insensitive) optical damping $C_0 \gamma$.
Alternatively, this cancellation could be viewed as being the result of a perfect cancellation of standard ``backaction" and ``imprecision" contributions to the added noise.  
%: This quadrature experiences two damping contributions, the paramp-modified intrinsic $\gamma + 2 \abs{\lambda}$ and the (phase-insensitive) optical damping $C_0 \gamma$, and the imprecision noise vanishes on-resonance when these rates are equal. 
The added noise away from resonance is shown in Fig.\@ \ref{fig:force_sensing}b. 

Note that when the impedance-matching condition is satisfied (so that $\gain = \abs{s_{YY} [0]}^2 = 0$), the mechanical input force quadrature $\hat{U}_{\mathrm{in}} [0]$ is transduced to $\hat{Y}_{\mathrm{out}} [0]$ with coefficient 
\begin{equation}
	s_{YU} [0] \big|_{\mathrm{imp. match}} = \frac{1}{\sqrt{1 - 2 \lambda / \gamma}}.
\end{equation}
Thus, if one tunes $\lambda$ to be slightly below $\gamma / 2$ while at the same time tuning $C_0$ to be $1 - 2 \lambda / \gamma$, our approach provides large-gain force detection with no added optical noise.  Note that by fixing $C_0 =  1 - 2 \lambda / \gamma$ to enforce impedance matching, the system hits instability at $\lambda = \gamma/2$.
Hence, the large force-detection gain in this regime (achieved with $C_0 \ll 1$) is directly associated with the expected amplification near the instability threshold.
%\BL{Note that when holding $C_0 + 2 \lambda / \gamma = 1$ fixed, $C_0$ approaches zero as $\lambda \rightarrow \gamma/2$, and the system approaches instability. Large force-detection gain is thus possible even though $C_0 \ll 1$ in this limit.}

When the two-phonon drive is off ($\lambda = 0$), one has instead
\begin{equation}
	s_{YU} [0] \big|_{\lambda = 0} = \frac{2 \sqrt{C_0}}{1 + C_0}.
\end{equation}
Note that this is never larger than unity --- without the mechanical parametric driving, one phonon's worth of input force produces at most one photon's worth of output light. While impedance matching is still possible (in this case by taking $C_0 = 1$), the lack of ``excitation-number gain" means that without parametric driving, the system provides only \textit{transduction} of the mechanical force, and not a true \textit{measurement} of the same.

%It is also possible to suppress the imprecision noise without meeting the impedance-matching condition. From the scattering matrix in Eq.~\eqref{eq:resonant_scattering}, one finds that in the large-$\gain$ limit,
%\begin{equation}
%	\bar{n}_{\mathrm{FD}} [0] = \frac{1}{2} \times \abs{\frac{s_{XX} [0]}{s_{XV} [0]}}^2 \approx \frac{1}{2} \times \left\lbrace \frac{C_0}{\gain} - 
%		\frac{2}{\sqrt{\gain}} + \frac{1}{C_0} \right\rbrace
%\end{equation}
%where we have assumed a zero-temperature cavity bath ($\nbarTc = 0$). The imprecision noise can thus be made to vanish by taking $\gain \gg C_0 \gg 1$. In this limit, the contributions to $\hat{X}_{\mathrm{out}} [0]$ from the optical and mechanical inputs are both suppressed; however, the optical contribution (i.e.\@ the imprecision noise) is suppressed more strongly.

The standard quantum limit on force-detection (force-detection SQL)
is  $\nAddFD [\omega] \ge 1/2$ (see e.g.\@ \cite{Caves:1980rmp, clerk_RMP, Clerk:2008bae, Hertzberg:2010bae}), and applies to any measurement that probes
 \textit{both} quadratures of a mechanical input force by monitoring \textit{both} position quadratures of a mechanical oscillator. Recent optomechanical experiments \cite{Schreppler:2014sql} have come close to reaching the force-detection SQL. We have seen how our scheme can be used to surpass the force-detection SQL for a single force quadrature by suppressing the optical noise floor in the on-resonance $\hat{Y}_{\mathrm{out}}$ quadrature while amplifying the mechanical response to the input force signal. This differs from backaction-evasion techniques (e.g.\@ \cite{Clerk:2008bae, Hertzberg:2010bae}), which surpass the force-detection SQL by producing a large signal without correspondingly raising the noise floor (but also without the suppression of that noise floor as afforded by our scheme). There also exist multi-mode approaches to sub-SQL force-detection \cite{Tsang:2010noisecancellation, Tsang:2012evadingQM, Woolley:2013twomodebae, Polzik:2015trajectories} --- by involving multiple resonators, these approaches can circumvent the force-detection SQL while providing detection of \textit{both} force quadratures.

We note that as mentioned above, the force-detection enhancement in this system relies on small cooperativity $C_0 < 1$; we assume that this is achieved by taking a sufficiently weak red-sideband drive such that $G \ll \kappa$, while still maintaining the good-cavity limit $\kappa \ll \omega_m$ and hence the validity of the RWA. We also assume that the drive is not \textit{too} weak, so that the single-photon optomechanical nonlinearity remains unimportant (i.e.\@ $G \gg g$).

%Note that the enhancement of force sensitivity afforded by our system is connected to the system's previously-discussed squeezing behaviour. While optimal squeezing and optimal force sensing do \textit{not} occur for the same choices of parameters, impedance matching (which optimizes force sensing) \textit{does} result in on-resonance squeezing provided that $C_0 > 2 \nbarTm + 1$:
%\begin{equation}
%	\bar{S}_{XX}^{\mathrm{out}} [0] \big|_{\mathrm{imp. match}} = \frac{\nbarTm + 1/2}{C_0}.
%\end{equation}

%\begin{figure}[t]
%	\includegraphics[width=0.5\textwidth]{force_imprecision_noise}
%	\caption{Imprecision noise for a measurement of the mechanical force quadrature $V_{\mathrm{in}} [\omega]$ (see \eqref{eq:detected_force_quadrature}), expressed as an equivalent number of mechanical bath noise quanta accompanying the force to be detected. The parametric gain is taken to be $\gain \equiv |s_{YY} [0]|^2 = 100$. By satisfying the impedance matching condition $C_0 = 1 + \sqrt{\gain}$, the imprecision noise can be made to vanish on-resonance. This plot takes $\kappa / \gamma = 10^5$, and assumes the RWA.}
%	\label{fig:force_imprecision_noise}
%\end{figure}

%%%%%%%%%%%%%%%%%%%%%%%%%%%%%%%%%%%%%%
\section{OMIT and negative spectral functions} \label{sec:SpectralFunction}

We now return to our system's intracavity properties, focusing on unusual features in the photonic dynamics.  These are most apparent in the behaviour of the cavity photon spectral function $A[\omega]$ (defined below), a quantity which usually plays the role of an effective density of states \cite{mahan}, but which can become negative here.  As we discuss, this indicates an effective negative temperature for the cavity photons at frequencies near resonance.  More concretely, it results in unusual behaviour in an OMIT-style experiment.

We imagine that in addition to the main input/output port through which the red-sideband laser drive is applied, the cavity is also coupled very weakly to a second waveguide at rate $\kappa^{\prime} \ll \kappa$. We will show that, surprisingly, near-resonant signals in this second waveguide can be reflected with gain even though the waveguide is severely undercoupled (and hence impedance mismatched).  We stress that such behaviour does not occur in typical resonantly-pumped quantum amplifiers, such as a standard DPA (see Appendix \ref{sec:DPA_comparison_appendix}).

As shown in Appendix \ref{sec:spectral_function_appendix}, the power reflection coefficient for such signals (averaged over their phase) is given by
\begin{equation}	\label{eq:spectral_function_reflection}
	\mathcal{R} [\omega] = 1 - \kappa^{\prime} A[\omega] + \mathcal{O} \left( (\kappa^{\prime})^2 \right).
\end{equation}
$A[\omega]$ is the cavity spectral function, defined as
\begin{equation}	\label{eq:define_spectral_function}
	A[\omega] = - 2 \mathrm{Im} \, G^{R} [\omega],
\end{equation}
where $G^R (t) = -i \theta (t) \langle [ \hat{d} (t), \hat{d}^{\dagger} (0) ] \rangle$ is the cavity retarded Green's function \footnote{We use the Fourier transform convention where $f (t) = \int_{- \infty}^{+\infty} \frac{\mathrm{d} \omega}{2 \pi} e^{-i \omega t} f [\omega]$ and $f [\omega] =  \int_{- \infty}^{+\infty} \mathrm{d} t \, e^{i \omega t} f(t)$.}.
$A[\omega]$ is usually interpreted as an effective density of single-particle states. As such, the familiar phenomenon of OMIT can be interpreted as an optomechanically-induced suppression of the density of photon states at the cavity resonance, i.e.\@ $A[\omega \sim 0] \rightarrow 0$: incident near-resonant photons in the weakly-coupled auxiliary waveguide don't see any available states when they reach the cavity, and are hence perfectly reflected ($\mathcal{R} \rightarrow 1$) \footnote{While this approach to OMIT may appear overly complicated, it allows for the examination of many non-standard situations such as those considered in \cite{Lemonde2013, Borkje:2013nonlinearsignatures, Kronwald:2013nonlinearomit}}.

\begin{figure}[t!]
	\includegraphics[width=0.45\textwidth]{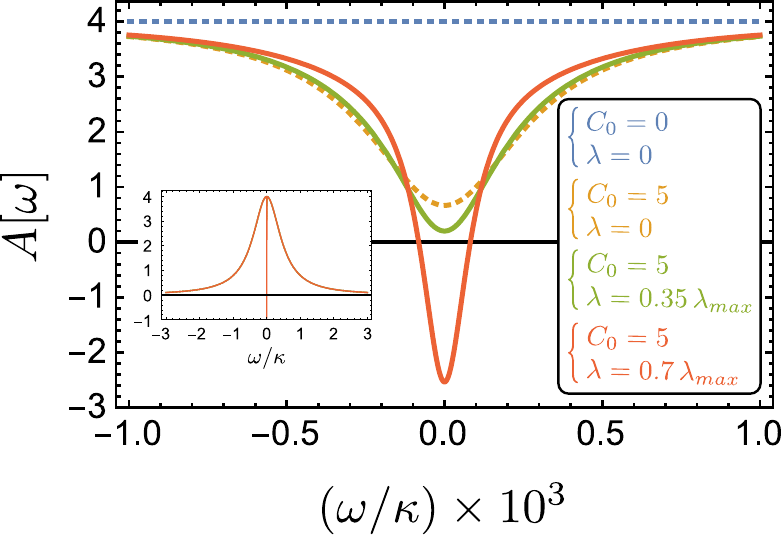}
	\caption{The cavity spectral function $A[\omega]$ (see Eq.~\eqref{eq:define_spectral_function}) can become negative near resonance, signalling the breakdown of its usual interpretation as a density of single-particle states. Without the optomechanical interaction ($C_0 = 0$), $A[\omega]$ follows a Lorentzian lineshape, decaying over the cavity linewidth $\kappa$ (see inset away from $\omega = 0$). Turning on the optomechanical interaction ($C_0 > 0$), ordinary OMIT physics reduces $A[\omega]$ near resonance (main plot) when the mechanical parametric drive is off (i.e.\@ $\lambda = 0$). When the parametric drive is sufficiently strong, the ``OMIT notch" in 
$A[\omega]$ passes through zero. We take $\gamma / \kappa = 10^{-4}$ for this plot, which assumes the RWA.}
	\label{fig:spectral_function}
\end{figure}

The above statements can of course be made more precise.  In the absence of parametric driving, one finds from Eq.~(\ref{eq:self_energy}) that on-resonance, the cavity self-energy is
\begin{equation}
	\Sigma_d [0] \big|_{\lambda = 0} = - \frac{2 i G^2}{\gamma},
\end{equation}
and so the spectral function is
\begin{multline}
	A[0] \big|_{\lambda = 0} = -2 \, \mathrm{Im} \left\lbrace \frac{1}{i \kappa/2 - \Sigma_d [0] \big|_{\lambda = 0}} \right\rbrace \\
		= \frac{4}{\kappa} \frac{1}{1 + C_0}.
\end{multline}
Increasing the cooperativity $C_0$ from $0$ effectively increases the damping felt by the cavity on-resonance, producing the familiar OMIT notch in $A[\omega]$. As mentioned, this notch can be interpreted as reflecting a lack of single-particle states near resonance (see the dashed orange curve in Fig.\@ \ref{fig:spectral_function}).  

%\BL{While this (weighted) density-of-states interpretation of $A[\omega]$ is intuitive, it is not always valid under non-equilibrium conditions. In particular, it is clear that a density of states cannot be negative.} 
Including now a non-zero parametric drive $\lambda$ in our system, we find something surprising:  increasing $\lambda$ from zero can increase the OMIT suppression of $A[\omega]$ near resonance, and can even push it below zero, making the spectral function negative
(see Fig.\@ \ref{fig:spectral_function}). Assuming as throughout that the strong-coupling regime is avoided, we find that $A[0]$ becomes negative when $2|\lambda| \geq \sqrt{1+ C_0} \gamma$.  As the system remains stable as long as $2 |\lambda| \leq (1+ C_0) \gamma$ (c.f. Eq.~(\ref{eq:stability})), in the large $C_0$ limit, there is a large parameter regime where the system is stable but exhibits a negative spectral function.

It immediately follows from Eq.~(\ref{eq:spectral_function_reflection}) that if $A[\omega] < 0$, then in an OMIT experiment, probe signals in an arbitrarily weakly coupled auxiliary waveguide (i.e.~$\kappa' \ll \kappa$) can be reflected with above-unity gain.  We stress that such stable negativity in the cavity spectral function does not occur in the standard OMIT setup, nor in a standard resonantly-pumped paramp.  
This is shown in Appendix \ref{sec:DPA_comparison_appendix}.  We also show in this appendix that our negative spectral function is directly connected to the effective negative cavity damping induced by the optomechanical interaction (as described by $- 2 \textrm{Im } \Sigma_d[\omega]$, c.f.~Eq.~(\ref{eq:self_energy})).
 
Note that it is of course possible to measure $A[\omega]$ without the need for an auxiliary waveguide; the cavity scattering matrix can be easily measured in an experiment, yielding the susceptibility (since $\matr{s}_{\mathrm{cav}} = \matr{1} - \kappa \matr{\chi}_{\mathrm{cav}}$), from which the spectral function can be extracted ($A = 2 \mathrm{Re} \, \chi_{dd}$). We present the previously-described experiment in order to emphasize the \textit{role} of $A[\omega]$.

Returning to the lab frame, $G^R(t)$ remains time-translation invariant, and the notch and negativity in $A[\omega]$ occurs at frequency $\omega$ near the cavity resonance frequency $\omega_{c}$.   For a time-independent Hamiltonian system in a time-independent state, $A [\omega > 0] < 0$ necessarily implies a stationary population inversion between eigenstates separated by $\hbar \omega$.
In our case, we have an open system and a time-dependent Hamiltonian.  Nonetheless,  the negativity in $A[\omega]$ is still indicative of population inversion.  This is best seen by computing the effective temperature of the cavity photons, a quantity that can be defined via the photonic noise properties.  As the system is not in equilibrium, this temperature will be explicitly frequency dependent (see Ref.~\cite{clerk_RMP} for an extensive, pedagogical discussion).  Formally, it is defined by comparing the size of the classical symmetrized photon correlation function (the so-called Keldysh Green function \cite{Kamenev}) to the size of the spectral function:
\begin{equation}
	\coth \left(   \frac{\omega}{2 T_{\mathrm{eff}}[\omega]} \right) \equiv
		\frac{-i G^K[\omega]}{A[\omega]} \equiv \frac{\int dt e^{i \omega t} \expec{ \lbrace \hat{d} (t), \hat{d}^{\dagger} (0) \rbrace}}{A[\omega]} 
\end{equation}
%\begin{equation}
%	G^K [\omega] =  i \coth \left(\frac{1}{2} \frac{\omega}{T_{\mathrm{eff}} [\omega]} \right) A [\omega].
%\end{equation}
In thermal equilibrium, $T_{\rm eff}[\omega]$ coincides with the system temperature $T$ at all frequencies.  Out of equilibrium, as the numerator on the RHS is always positive definite, a negative spectral function at $\omega > 0$ necessarily implies a negative temperature at that frequency.

We stress that the effective temperature $T_{\rm eff}[\omega]$ also has a direct operational meaning, which we elucidate by considering another different experiment.
As discussed extensively in Ref.~\cite{clerk_RMP}, if one were to weakly couple a qubit with a splitting frequency $\Omega = \omega_{c}$ to the cavity photons via an interaction Hamiltonian $H_{\rm int} \propto \left(\hat{\sigma}_+ \hat{d} + h.c.\right)$, then the cavity photons would act as a bath for the qubit. The corresponding steady state of the qubit would correspond to a thermal state at temperature $T_{\rm eff}[\omega_{c}]$:
\begin{equation}
	\expec{\hat{\sigma}_z} = - \tanh \left( \frac{\hbar \omega_c}{2 k_B T_{\mathrm{eff}} [\omega_c]} \right).
\end{equation} 
Hence, a negative effective cavity temperature would directly translate into a simple population inversion of the qubit.

\begin{figure}[t]
	\includegraphics[width=\columnwidth]{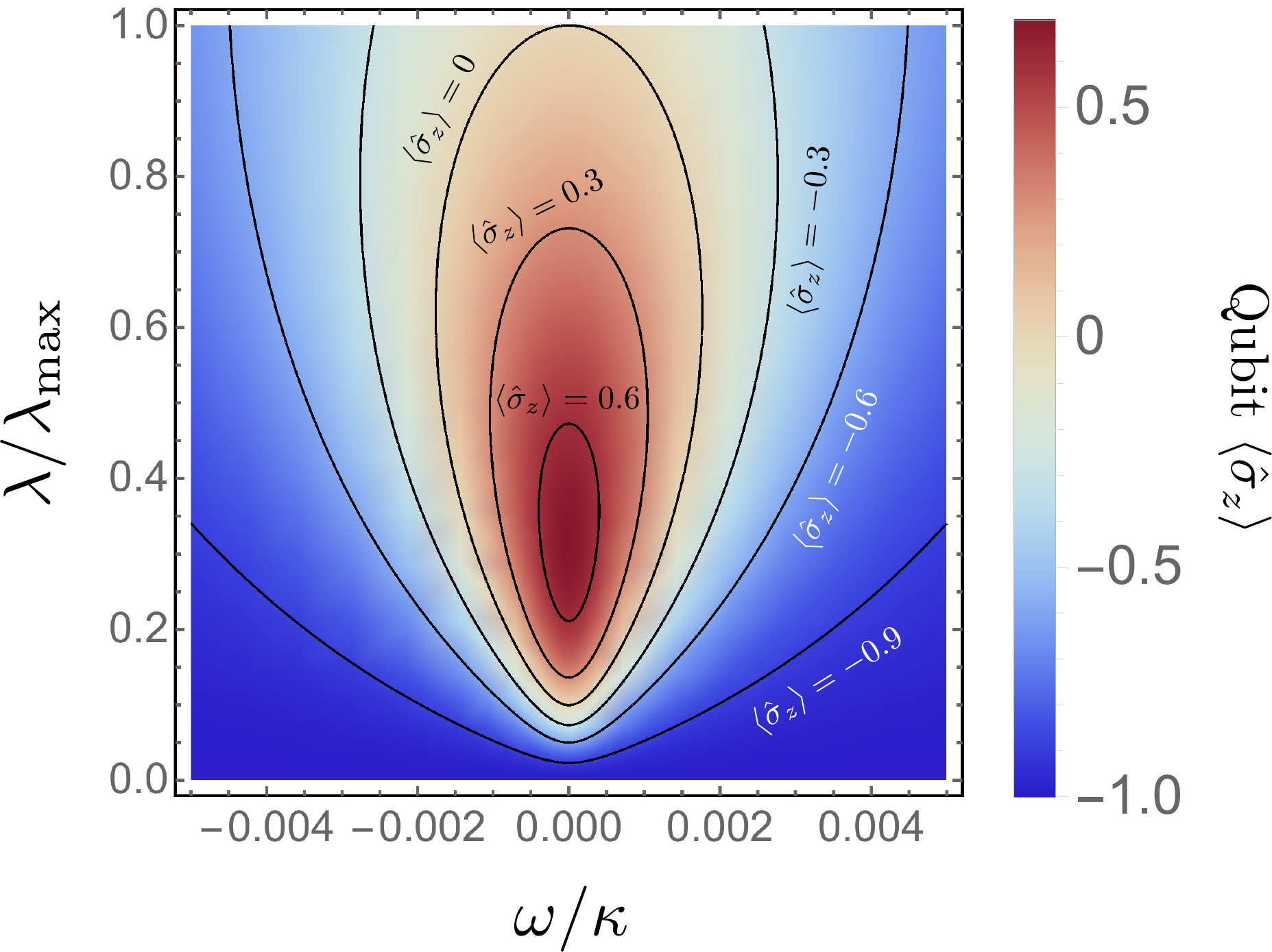}
	\caption{Polarization $\expec{\hat{\sigma}_z}$ of a qubit (level splitting $\Omega = \omega + \omega_c$) weakly coupled to the cavity mode via $\hat{H}_{\mathrm{int}} \propto (\hat{\sigma}_+ \hat{d} + h.c.)$. The qubit polarization is precisely the same as if it were coupled to a thermal reservoir at temperature $T_{\mathrm{eff}} [\Omega]$; the regions where $\langle \hat{\sigma}_z \rangle > 0$ directly indicate a negative effective photon temperature. This plot takes $\gamma/\kappa = 10^{-4}$ and $C_0 = 100$, and assumes the RWA.}
	\label{fig:qubit_inversion}
\end{figure}

One can show that for such a setup, the maximum qubit polarization $\expec{\hat{\sigma}_z}$ occurs when $2 | \lambda | / \gamma = (1+C_0)^{3/4}$. For large $C_0$, this maximum is
\begin{equation}
	\expec{\hat{\sigma}_z} \big|_{\mathrm{max}} = 1 - \frac{4}{\sqrt{C_0}} + \mathcal{O} \left( \frac{1}{C_0} \right).
\end{equation}
We therefore find that in the $C_0 \rightarrow \infty$ limit, the qubit becomes completely inverted, i.e.\@ the effective photon temperature becomes infinitesimally negative ($\beta_{\mathrm{eff}} [\omega_c] \rightarrow - \infty$). One could also imagine a similar experiment with a qubit detuned from $\omega_c$ to probe the effective photon temperature at other frequencies --- see Fig.~\ref{fig:qubit_inversion}.

While we can rigorously associate a negative temperature to our system (in the lab frame), a further discussion of the relevant population inversion is difficult.  In our interaction picture, the Hamiltonian and steady-state are time-independent, and the negative spectral function indicates an anomalous population of the system-plus-bath energy eigenstates.  However, in the lab frame, these states do not correspond to energy eigenstates or even Floquet eigenstates (as in general, the mechanical and cavity frequencies are incommensurate, so the lab-frame Hamiltonian is not periodic).  This being said, the negative spectral function and negative effective temperature implies that for any weak, single-photon probes of the cavity, it effectively behaves like a {\it time-independent} system with a conventional population inversion.

It is interesting to note that the basic mechanism in our system which allows a stable negative photonic spectral function can be generalized to other more complex systems.  One can begin with any parametrically-driven unstable mode. If this mode is then stabilized by coupling to a damped auxiliary mode, there can exist a range of parameters where the auxiliary mode displays a negative spectral function.

\section{Conclusion}
We have described a simple twist on the standard optomechanical setup which can be used to translate mechanical degenerate parametric driving into squeezing and amplification of an optical mode. We have shown how our system can approach the quantum limit when operated as a phase-sensitive amplifier, and can produce significant degrees of output squeezing when operated as a squeezer, all while avoiding the need for a conventional optical nonlinearity. We have highlighted the differences between our method and other previously-described optomechanical amplification and squeezing protocols. We have shown how our method can yield single-quadrature force measurement beyond the force-detection standard quantum limit. Finally, we have found that this method leads to an unusual situation involving a negative cavity spectral function, and have briefly discussed the implications of this negativity.  

We thank Jack Sankey for useful conversations.  This work was supported by NSERC.

%%%%%%%%%%%%%%%%%%%%%%%
\appendix
%\begin{widetext}
\section{Susceptibility}	\label{sec:susc_appendix}
Using input-output theory to deal with the dissipative environment, the RWA Hamiltonian $\hat{H} = G \left( \hat{d}^{\dagger}\hat{b} + \hat{b}^{\dagger} \hat{d} \right) + \frac{i }{2} \left( \lambda \hat{b^{\dagger}} \hat{b}^{\dagger} - \lambda^* \hat{b} \hat{b} \right)$  yields Heisenberg-Langevin equations with solution
\begin{widetext}
\begin{equation}
	\matr{Q} [\omega] = 
		\begin{pmatrix}
			\hat{X} [\omega] \\
			\hat{Y} [\omega]] \\
			\hat{U} [\omega] \\
			\hat{V} [\omega]
		\end{pmatrix}
		= - \boldsymbol{\chi} [\omega] \left\lbrace
		\begin{pmatrix}
			\sqrt{\kappa_{\mathrm{ex}}} & 0 & 0  & 0 \\
			0 & \sqrt{\kappa_{\mathrm{ex}}} & 0 & 0 \\
			0 & 0 & \sqrt{\gamma} & 0 \\
			0 & 0 & 0 & \sqrt{\gamma}
		\end{pmatrix} \matr{Q}_{\mathrm{in}} [\omega]
		+ \begin{pmatrix}
			\sqrt{\kappa_{\mathrm{int}}} & 0 & 0  & 0 \\
			0 & \sqrt{\kappa_{\mathrm{int}}} & 0 & 0 \\
			0 & 0 & 0 & 0 \\
			0 & 0 & 0 & 0
		\end{pmatrix} \matr{Q}_{\xi}^{\mathrm{(int)}} [\omega]
	\right\rbrace.
\end{equation}
\end{widetext}
$\matr{Q}_{\xi}^{\mathrm{(int)}}  [\omega]$ consists of the quadratures of the noise operators corresponding to the internal loss port ($\kappa_{\mathrm{int}}$), while $\matr{Q}_{\mathrm{in}} [\omega]$ contains the quadratures of the noise operators correponding to the signal port ($\kappa_{\mathrm{ex}}$) and mechanical noise port ($\gamma$). The total cavity damping is $\kappa = \kappa_{\mathrm{int}} + \kappa_{\mathrm{ex}}$. The susceptibility $\boldsymbol{\matr{\chi}} [\omega]$ is given by
\begin{widetext}
\begin{equation} \label{eq:susceptibility}
	\boldsymbol{\chi} [\omega] =
	\begin{pmatrix}
		\frac{1}{\chi_o^{-1}[\omega]  + G^2 \chi_{m,+}[\omega]} 
		& 0 & 0 & \frac{G}{\chi_o^{-1}[\omega] \chi_{m,+}^{-1}[\omega] + G^2} \\
		0 & \frac{1}{\chi_o^{-1}[\omega] + G^2 \chi_{m,-}[\omega]}
		& - \frac{G}{\chi_o^{-1}[\omega] \chi_{m,-}^{-1}[\omega] + G^2} & 0 \\
		0 & \frac{G}{\chi_o^{-1}[\omega] \chi_{m,-}^{-1}[\omega] + G^2}
		& \frac{1}{\chi_{m,-}^{-1}[\omega] + G^2 \chi_o [\omega]} & 0 \\
		- \frac{G}{\chi_o^{-1}[\omega] \chi_{m,+}^{-1}[\omega] + G^2} & 0 & 0 
		& \frac{1}{\chi_{m,+}^{-1}[\omega] + G^2 \chi_o[\omega] } 
	\end{pmatrix}
\end{equation}
\end{widetext}
where we have defined $\chi_o^{-1} [\omega] = -i \omega + \kappa / 2$ and $\chi_{m,\pm}^{-1} [\omega] = - i \omega + \gamma / 2 \pm \abs{\lambda}$. $\boldsymbol{\chi} [\omega]$ is precisely the susceptibility for a red-detuned optomechanical cavity in the RWA with the mechanical damping $\gamma$ modified by the parametric drive in the usual phase-sensitive way: In terms involving $\hat{U}$ (and, by extension, the cavity quadrature $\hat{Y}$ which it couples to), one has $\gamma \rightarrow \gamma - 2 \abs{\lambda}$, while for the terms involving $\hat{V}$ (and hence $\hat{X}$), the replacement is $\gamma \rightarrow \gamma + 2 \abs{\lambda}$.

\section{Stability and Mode-Splitting}	\label{sec:stability_appendix}
The stability of the system can be determined from the poles of the susceptibility matrix $\matr{\chi}[\omega]$ (see Eq.~\eqref{eq:susceptibility}). Two of these poles lie at
\begin{equation}	\label{eq:plus_poles}
	\Omega^{(+)}_{\pm} = - \frac{i}{2} \left( \frac{\kappa}{2} + \frac{\gamma}{2} + \abs{\lambda}
		\pm \sqrt{ \left( \frac{\kappa}{2} - \frac{\gamma}{2} - \abs{\lambda} \right)^2 - 4G^2 } \right)
\end{equation}
and the other two lie at 
\begin{equation}	\label{eq:minus_poles}
	\Omega^{(-)}_{\pm} = - \frac{i}{2} \left( \frac{\kappa}{2} + \frac{\gamma}{2} - \abs{\lambda}
		\pm \sqrt{ \left( \frac{\kappa}{2} - \frac{\gamma}{2} + \abs{\lambda} \right)^2 - 4G^2 } \right).
\end{equation}
Maintaining stability requries that these poles lie in the lower half-plane, and avoiding a mode-splitting requires that they lie on the imaginary axis. The poles at $\Omega^{(+)}_{\pm}$ will always lie in the lower half-plane, as these are the same as in the case of a red-sideband-driven linearized optomechanical cavity with a modified but still positive mechanical damping rate; such a system is always stable. Keeping $\Omega^{(-)}_{-}$ in the lower half-plane is thus sufficient to maintain stability, and assuming that mode-splitting is avoided, this is equivalent to the condition Eq.~\eqref{eq:stability}, i.e.\@ $\abs{\lambda} < (\gamma / 2) (1 + C_0)$. If there is a mode-splitting, i.e.\@ if the square-root in $\Omega^{(+/-)}_{\pm}$ is imaginary, then stability requires $\abs{\lambda} < \frac{\kappa + \gamma}{2}$. Taking these together yields
\begin{equation}
	\abs{\lambda} < \min{ \left\lbrace \frac{\kappa + \gamma}{2}, \frac{\gamma}{2}
		\left( 1 + C_0 \right) \right\rbrace}.
\end{equation}

Mode-splitting is avoided if the square-root in Eq.~\eqref{eq:plus_poles} is real, i.e.\@ if
\begin{equation}
	\left( \frac{\kappa}{2} - \frac{\gamma}{2} - \abs{\lambda} \right)^2 >  4G^2.
\end{equation}
Using the relevant stability condition, we see that this avoidance is achieved over all stable values of $\lambda$ provided that
\begin{equation}
	\left( 1 - \frac{2 \gamma}{\kappa} - \frac{4G^2}{\kappa^2} \right)^2 > \frac{16G^2}{\kappa^2}.
\end{equation}
Because $\gamma \ll \kappa$, it then follows that weak coupling ($4G^2 / \kappa^2 \ll 1$) is sufficient to avoid mode-splitting.

\section{Scattering}	\label{sec:scattering_appendix}
Input-output theory leads to the following simple expression for the radiation leaving the cavity:
\begin{equation}
	\hat{d}_{\mathrm{out}} = \hat{d}_{\mathrm{in}} + \sqrt{\kappa_{ex}} \hat{d}.
\end{equation}
Going over to the quadrature basis, we can thus write
\begin{equation}
	\hat{\matr{Q}}_{\mathrm{out}} [\omega] = \matr{s} [\omega] \hat{\matr{Q}}_{\mathrm{in}} [\omega]
		 + \matr{N} [\omega] \hat{\matr{Q}}_{\xi}^{\mathrm{(int)}}  [\omega].
\end{equation}
with the matrix 
\begin{equation}
	\matr{s}[\omega] = 
	\matr{1} -
	\begin{pmatrix}
		\sqrt{\kappa_{ex}} & 0 & 0 & 0 \\
		0 & \sqrt{\kappa_{ex}} & 0 & 0 \\
		0 & 0 & \sqrt{\gamma} & 0 \\
		0 & 0 & 0 & \sqrt{\gamma}
	\end{pmatrix}
	\boldsymbol{\chi} [\omega]
	\begin{pmatrix}
		\sqrt{\kappa_{ex}} & 0 & 0 & 0 \\
		0 & \sqrt{\kappa_{ex}} & 0 & 0 \\
		0 & 0 & \sqrt{\gamma} & 0 \\
		0 & 0 & 0 & \sqrt{\gamma}
	\end{pmatrix}	
\end{equation}
describing the scattering of an incident signal, and with the matrix
\begin{equation}
	\matr{N}[\omega] = 	-
	\begin{pmatrix}
		\sqrt{\kappa_{ex}} & 0 & 0 & 0 \\
		0 & \sqrt{\kappa_{ex}} & 0 & 0 \\
		0 & 0 & \sqrt{\gamma} & 0 \\
		0 & 0 & 0 & \sqrt{\gamma}
	\end{pmatrix}
	\boldsymbol{\chi} [\omega]
	\begin{pmatrix}
		\sqrt{\kappa_{int}} & 0 & 0 & 0 \\
		0 & \sqrt{\kappa_{int}} & 0 & 0 \\
		0 & 0 & 0 & 0 \\
		0 & 0 & 0 & 0
	\end{pmatrix}	
\end{equation}
bringing in the noise associated with internal loss in the cavity. 
%If there are internal losses ($\kappa_{\mathrm{int}} \ne 0$), the scattering matrix is 
%\begin{equation}
%	\matr{s} [\omega] = \\
%	\begin{pmatrix}
%		1 - \frac{\kappa_{\mathrm{ex}}}{\chi_o^{-1}[\omega]  + G^2 \chi_{m,+}[\omega]} & 0 & 0
%		& \frac{- G \sqrt{\kappa_{\mathrm{ex}} \gamma}}
%			{\chi_o^{-1}[\omega] \chi_{m,+}^{-1}[\omega] + G^2} \\
%		0 & 1 - \frac{\kappa_{\mathrm{ex}}}{\chi_o^{-1}[\omega]  + G^2 \chi_{m,-}[\omega]}
%		&  \frac{G \sqrt{\kappa_{\mathrm{ex}} \gamma}}
%			{\chi_o^{-1}[\omega] \chi_{m,-}^{-1}[\omega] + G^2} & 0 \\
%		0 &  \frac{- G \sqrt{\kappa_{\mathrm{ex}} \gamma}}
%			{\chi_o^{-1}[\omega] \chi_{m,-}^{-1}[\omega] + G^2}
%		& 1 - \frac{\gamma}{\chi_{m,-}^{-1}[\omega]  + G^2 \chi_o[\omega]} & 0 \\
%		\frac{G \sqrt{\kappa_{\mathrm{ex}} \gamma}}
%			{\chi_o^{-1}[\omega] \chi_{m,+}^{-1}[\omega] + G^2} & 0 & 0
%		&  1 - \frac{\gamma}{\chi_{m,+}^{-1}[\omega]  + G^2 \chi_o[\omega]}
%	\end{pmatrix}
%\end{equation}
%and the extra noise is brought in by
%\begin{equation}
%	\matr{N} [\omega] =
%	\begin{pmatrix}
%		- \frac{ \sqrt{ \kappa_{\mathrm{ex}} \kappa_{\mathrm{int}}}}
%			{\chi_o^{-1} [\omega] + G^2 \chi_{m,+} [\omega]} & 0 & 0 & 0 \\
%		0 & - \frac{ \sqrt{ \kappa_{\mathrm{ex}} \kappa_{\mathrm{int}}}}
%			{\chi_o^{-1} [\omega] + G^2 \chi_{m,-} [\omega]} & 0 & 0 \\
%		0 & - \frac{G \sqrt{ \kappa_{\mathrm{int}} \gamma}}
%			{\chi_o^{-1}[\omega] \chi_{m,-}^{-1}[\omega] + G^2} & 0 & 0 \\
%		\frac{G \sqrt{ \kappa_{\mathrm{int}} \gamma}}
%			{\chi_o^{-1}[\omega] \chi_{m,+}^{-1}[\omega] + G^2} & 0 & 0 & 0
%	\end{pmatrix}.
%\end{equation}

In the case where there are no internal losses, i.e.\@ $\kappa_{\mathrm{int}} = 0, \kappa = \kappa_{\mathrm{ext}}$, one finds
\begin{widetext}
\begin{multline}	\label{eq:freq_dependent_scattering}
	 \matr{s} [\omega] = \\
	\begin{pmatrix}
			\frac{G^2 + (-i \omega - \kappa / 2)(-i \omega + \gamma / 2 + \abs{\lambda})}
				{G^2 + (-i \omega + \kappa / 2)(-i \omega + \gamma / 2 + \abs{\lambda})} & 0 & 0
			& \frac{- G \sqrt{\kappa \gamma}}{G^2 + (-i \omega + \kappa / 2)
				(-i \omega + \gamma / 2 + \abs{\lambda})}  \\
			0 &  \frac{G^2 + (-i \omega - \kappa / 2)(-i \omega + \gamma/ 2 - \abs{\lambda})}
				{G^2 + (-i \omega + \kappa / 2)(-i \omega + \gamma / 2 - \abs{\lambda})}
			& \frac{G \sqrt{\kappa \gamma}}{G^2 + (-i \omega + \kappa / 2)(-i \omega + \gamma / 2 - \abs{\lambda})}
			& 0 \\
			0 & \frac{-G \sqrt{\kappa \gamma}}{G^2 + (-i \omega + \kappa / 2)
				(-i \omega + \gamma / 2 - \lambda)}
			& \frac{G^2 + (-i \omega + \kappa / 2)(- i\omega - \gamma / 2 - \abs{\lambda})}{
				G^2 + (-i \omega + \kappa / 2)(-i \omega + \gamma / 2 - \abs{\lambda})} & 0 \\
			 \frac{G \sqrt{\kappa \gamma}}{G^2 + (-i \omega + \kappa / 2)
				(-i \omega + \gamma / 2 + \abs{\lambda})} & 0 & 0
			& \frac{G^2 + (-i \omega + \kappa / 2)(- i\omega - \gamma / 2 + \abs{\lambda})}
				{G^2 + (-i \omega + \kappa / 2)(-i \omega + \gamma/ 2 + \abs{\lambda})} 
	\end{pmatrix}.
\end{multline}
\end{widetext}

\section{Bandwidth}	\label{sec:bandwidth_appendix}
To determine the amplifier bandwidth, consider the denominator of the $Y$-$Y$ scattering element:
\begin{align}
	D & = G^2 + \left( -i \omega + \frac{\kappa}{2} \right) \left( -i \omega + \frac{\gamma}{2} - \abs{\lambda} \right)
	\nonumber \\
		& = - \omega^2 - i \omega \left( \frac{\kappa}{2} + \frac{\gamma}{2} - \abs{\lambda} \right) + \frac{\kappa}{2}
			\left( \frac{\gamma}{2} - \abs{\lambda} \right) + G^2.
\end{align}
Approximating $s_{YY} [\omega]$ as Lorentzian, this gives a full-width at half-maximum (FWHM) of
\begin{equation}
	D = \frac{4G^2 + \kappa \left( \gamma - 2 \abs{\lambda} \right)}{\kappa + \gamma - 2 \abs{\lambda}}
		= \frac{8 G^2 \kappa}{4G^2 \left( 1 - \sqrt{\mathcal{G}} \right) + \kappa^2 \left( 1 + \sqrt{\mathcal{G}} \right)}.
\end{equation}
For large gain, this is well-approximated by
\begin{equation}
	D \approx \frac{1}{\sqrt{\mathcal{G}}} \frac{2 \kappa}{\left( \frac{\kappa}{2G} \right)^2 - 1}.
\end{equation}
Taking $\kappa \gg 2G$, we can further approximate
\begin{equation}
	D \sqrt{\mathcal{G}} \approx \frac{8 G^2}{\kappa}
\end{equation}
which expresses the gain-bandwidth limitation of our system.

\section{Beyond the RWA}	\label{sec:beyond_RWA_appendix}
The linearized optomechanical interaction involves beamsplitter terms ($\propto \hat{d} \hat{b}^{\dagger} + h.c.$) and entangling terms ($\propto \hat{d} \hat{b} + h.c$). When driving on the red sideband in the good-cavity limit, the entangling terms describe highly off-resonant processes and correspondingly oscillate rapidly in the interaction picture. Discarding these terms constitutes the rotating wave approximation (RWA). Without the RWA, the linearized Hamiltonian for our system is (in the interaction picture)
\begin{widetext}
\begin{equation}	\label{eq:full_Hamiltonian}
	\hat{H} = G \left( \hat{d}^{\dagger}\hat{b} + \hat{b}^{\dagger} \hat{d} \right)
		+ \frac{i }{2} \left( \lambda \hat{b}^{\dagger} \hat{b}^{\dagger} - \lambda^* \hat{b} \hat{b} \right)
		+ G \left( e^{- 2 i \omega_M t} \hat{d} \hat{b} 
		+ e^{2 i \omega_M t} \hat{d}^{\dagger} \hat{b}^{\dagger} \right)
\end{equation}
\end{widetext}
Including the counter-rotating terms makes the equations of motion dependent on time, coupling frequency components separated by $\pm 2 \omega_M$. We handle this complication by following a sideband truncation approach similar to \cite{Malz:2016floquet}, and focus on the stationary part of the noise. Similar techniques were used in \cite{Metelmann:2014dissipativeamp, Weinstein:2014asymmetry}. In the frequency domain,
\begin{subequations} \label{eq:nonRWAfourierHL}
	\begin{align}
		\left( -i \omega + \frac{\kappa}{2} \right) \hat{d} [\omega] + i G \hat{b} [\omega] 
			+ i G \hat{b}^{\dagger}\left[ \omega + 2 \omega_M \right] \nonumber \\
		= - \sqrt{\kappa_{\mathrm{ex}}} \hat{d}_{\mathrm{in}} [ \omega ] 
			- \sqrt{\kappa_{\mathrm{int}}} \hat{\xi} [\omega]
	\end{align}
	\begin{align}
		\left( -i \omega + \frac{\kappa}{2} \right) \hat{d}^{\dagger} [\omega]
			- i G \hat{b}^{\dagger} [\omega] - i G \hat{b} \left[ \omega - 2 \omega_M \right] 
			\nonumber \\
		= - \sqrt{\kappa} \hat{d}^{\dagger}_{\mathrm{in}} [ \omega ]
			- \sqrt{\kappa_{\mathrm{int}}} \hat{\xi}^{\dagger} [\omega]
	\end{align}
	\begin{align}
		\left( -i \omega + \frac{\gamma}{2} \right) \hat{b} [\omega] - \lambda \hat{b}^{\dagger} [\omega] 
			+ i G \hat{d} [\omega] + i G \hat{d}^{\dagger}\left[ \omega + 2 \omega_M \right]
			\nonumber \\
		 = - \sqrt{\gamma} \hat{b}_{\mathrm{in}} [ \omega ]
	\end{align}
	\begin{align}
		\left( -i \omega + \frac{\gamma}{2} \right) \hat{b}^{\dagger} [\omega] - \lambda^* \hat{b} [\omega] 
			- i G \hat{d}^{\dagger} [\omega] - i G \hat{d} \left[ \omega - 2 \omega_M \right]
			\nonumber \\
		 = - \sqrt{\gamma} \hat{b}^{\dagger}_{\mathrm{in}} [ \omega ].
	\end{align}
\end{subequations}
By shifting $\omega \rightarrow \omega \pm 2 \omega_M$ and substituting the resulting equations back into Eq.~\eqref{eq:nonRWAfourierHL}, one obtains an additional eight equations now involving operators evaluated at $\omega$, $\omega \pm 2 \omega_M$ and $\omega \pm 4 \omega_M$. Because we are interested in the behaviour near resonance and in the good-cavity limit, the response of the cavity is miniscule at the second-order sideband at $\omega \pm 4 \omega_M$, so we drop terms evaluated at these frequencies to close the set of 12 equations. If a better approximation is needed, one can instead iterate the shifting of $\omega$ by $\pm 2 \omega_M$ and include as many sidebands as desired.

\section{Comparison to DPA}	\label{sec:DPA_comparison_appendix}

\subsection{Resonant parametric amplifiers}		\label{subsec:resonant_paramps}
As discussed in the main text, our system bears a degree of resemblance to a true optical DPA. In this section we enable this comparison by recalling several properties of the DPA, and of parametric amplifiers in general.

The resonant \textit{non}-degenerate paramp involves two modes $\hat{a}_S$ and $\hat{a}_I$, and is pumped at $\omega_S + \omega_I$. In the interaction picture,
\begin{equation}
	\hat{H}_{\mathrm{NDPA}} = i (\mu \hat{a}_S^{\dagger} \hat{a}_I^{\dagger} - \mu^* \hat{a}_S \hat{a}_I).
\end{equation} 
For the degenerate paramp, $\hat{a}_S = \hat{a}_I \equiv \hat{a}$, $\omega_S = \omega_I \equiv \omega_c$, and $\mu \rightarrow \Lambda / 2$. The coherent DPA Hamiltonian is
\begin{equation}
	\hat{H}_{\mathrm{DPA}} = \frac{i}{2} \left( \Lambda \hat{a}^{\dagger} \hat{a}^{\dagger} - \Lambda^* \hat{a} \hat{a} \right).
\end{equation}

Dealing with coherent driving and dissipation via input-output theory, one obtains the equations of motion
\begin{subequations}		\label{eq:NDPA_eom}
	\begin{equation}	
		-i \omega \hat{a}_S [\omega] = - \frac{\kappa_S}{2} \hat{a}_S [\omega] + \mu \hat{a}_I^{\dagger}  [\omega]
			- \sqrt{\kappa_S} \hat{a}_{S, \mathrm{in}} [\omega]
	\end{equation}
	\begin{equation}	
		-i \omega \hat{a}_I^{\dagger} [\omega] = - \frac{\kappa_I}{2} \hat{a}_I^{\dagger} [\omega] + \mu \hat{a}_S [\omega]
			- \sqrt{\kappa_I} \hat{a}^{\dagger}_{I, \mathrm{in}} [\omega]
	\end{equation}
\end{subequations}
for the NDPA, and
\begin{subequations}		\label{eq:DPA_eom}
	\begin{equation}	
		-i \omega \hat{a} [\omega] = - \frac{\kappa}{2} \hat{a} [\omega] + \Lambda \hat{a}^{\dagger}  [\omega]
			- \sqrt{\kappa} \hat{a}_{\mathrm{in}} [\omega]
	\end{equation}
	\begin{equation}	
		-i \omega \hat{a}^{\dagger} [\omega] = - \frac{\kappa}{2} \hat{a}^{\dagger} [\omega] + \Lambda^* \hat{a} [\omega]
			- \sqrt{\kappa} \hat{a}^{\dagger}_{\mathrm{in}} [\omega]
	\end{equation}
\end{subequations}
for the DPA.
Comparing to Eqs.~\eqref{eq:cavity_EOM} and \eqref{eq:EOM_terms} in the main text, we see that our system resembles a DPA but with a frequency-dependent effective parametric drive strength $\Lambda \rightarrow \tilde{\lambda} [\omega]$ and with a non-zero cavity self-energy $\Sigma_d [\omega]$.

We will treat the NDPA case explicitly, and obtain results for the DPA by the simple replacements $\mu \rightarrow \Lambda$ and $\kappa_S, \kappa_I \rightarrow \kappa$. The equations of motion \eqref{eq:NDPA_eom} lead to the parametric amplifier susceptibility (in the field operator basis $( \hat{a}_S, \hat{a}_I^{\dagger} )^T$)
\begin{widetext}
\begin{equation}
	\matr{\chi}^{(\mathrm{paramp})} [\omega] = 
		\frac{1}{\left( -i \omega + \kappa_S / 2 \right) \left(-i \omega + \kappa_I / 2 \right) - | \mu |^2}
		\begin{pmatrix}
			-i \omega + \kappa_I / 2 & \mu \\
			\mu^* & -i \omega + \kappa_S / 2
		\end{pmatrix}.
\end{equation}
\end{widetext}
Stability for the resonant NDPA requires that
\begin{equation}	\label{eq:NDPA_stability}
	| \mu |^2 < \frac{\kappa_S \kappa_I}{4},
\end{equation}
reducing to the familiar condition $| \Lambda |^2 < \kappa^2 / 4$ in the resonant DPA case.

Recalling that $\chi_{a a} [\omega] = i G^R [\omega]$ where $G^R$ is the retarded Green's function for the mode $\hat{a}$, we find the spectral function to be
\begin{equation}	\label{eq:A_paramp}
	A^{(\mathrm{paramp})}_{S/I} [\omega] = \frac{\kappa_{I/S} \left( \frac{\kappa_S \kappa_I}{4} - | \mu |^2 \right) + \kappa_{S/I} \omega^2}					{\left( \frac{\kappa_S 	\kappa_I}{4} - |\mu|^2 - \omega^2 \right)^2 + \frac{\omega^2 \left(\kappa_S + \kappa_I \right)^2}{4}} 
	\ge 0
\end{equation}
where we have used the NDPA stability condition \eqref{eq:NDPA_stability} to obtain the final inequality $A^{(\mathrm{paramp})} [\omega] \ge 0$. This highlights that $A[\omega] < 0$ as observed in our system is not simply a generic non-equilibrium effect, and it does not occur in the simple resonant paramp (either degenerate or non-degenerate).

\subsection{Coherent vs.~dissipative couplings}
The parametric interaction in a (N)DPA is coherent, i.e.\@ it is completely Hamiltonian. This is reflected in the equation of motion Eq.~\eqref{eq:DPA_eom} by the fact that the terms where $\hat{a}$ drives $\hat{a}^{\dagger}$ and vice versa have coefficients which are complex conjugates of each other --- $\Lambda^* = (\Lambda)^*$. In our system, this is not quite the case. Instead, we have from Eqs.~\eqref{eq:cavity_EOM} and \eqref{eq:lambda_tilde} that $\hat{d}^{\dagger} [\omega]$ appears on the RHS of $\hat{d}$'s $\omega$-domain equation of motion with coefficient
\begin{equation}
	\tilde{\lambda} [\omega] =  \frac{G^2 \lambda}{(-i \omega + \gamma / 2)^2 - \abs{\lambda}^2},
\end{equation}
while $\hat{d} [\omega]$ appears on the corresponding equation for $\hat{d}^{\dagger} [\omega]$ with coefficient $(\tilde{\lambda} [-\omega])^*$ (recall that $(\hat{a} [\omega])^{\dagger} = \hat{a}^{\dagger} [- \omega]$). So, $\tilde{\lambda}$ can be thought of as representing a coherent effective interaction when $(\tilde{\lambda} [\omega])^* = (\tilde{\lambda} [-\omega])^*$, i.e.\@ when $\tilde{\lambda} [\omega] = \tilde{\lambda} [- \omega]$. Since $(\tilde{\lambda} [\omega])^* = \tilde{\lambda} [- \omega]$, we can think of $\mathrm{Re} \tilde{\lambda}$ as the ``coherent part" of the effective parametric interaction, and $\mathrm{Im} \tilde{\lambda}$ as the ``dissipative part." One finds that the ratio of the two is
\begin{equation}
	\left| \frac{\mathrm{Im} \tilde{\lambda}} {\mathrm{Re} \tilde{\lambda}} \right| = \frac{\abs{\omega} \gamma}{\abs{ \gamma^2 / 4 - \abs{\lambda}^2 - \omega^2 }}.
\end{equation}
We found our system to be most useful when the cooperativity $C_0$ is large, and significant amplification and squeezing set in when $\lambda$ approaches $(\gamma/2)(1+ C_0)$. Therefore, with practical parameter choices, one has $|\lambda| \gg \gamma$, and so 
\begin{equation}
	\left| \frac{\mathrm{Im} \tilde{\lambda}} {\mathrm{Re} \tilde{\lambda}} \right| \approx \frac{|\omega| \gamma}{\abs{ \abs{\lambda}^2 + \omega^2 }}.
\end{equation}
This means that with realistic parameter choices, the interaction is almost entirely coherent. For there to be any frequency where the dissipative component is significant, one must have $\abs{\lambda}^2 < \gamma^2 / 4$ (and then the dissipative component will dominate only when $\omega^2 - \gamma^2 / 4 + \abs{\lambda}^2 \ll |\omega| \gamma$).

We note that a dissipative NDPA interaction has been considered previously in other works \cite{Buchmann:2013conjugation, Metelmann:2014dissipativeamp}. Like its coherent counterpart, the spectral function of a resonant dissipative NDPA remains positive at all frequencies.

\subsection{Understanding $A<0$: Mapping to the DPA}	\label{subsec:negative_A_mapping}

We have seen previously how the effective cavity dynamics of our system described by Eqs.~\eqref{eq:cavity_EOM} and \eqref{eq:EOM_terms} resembles the dynamics of a DPA with self-energy $\Sigma_d [\omega]$ and effective (frequency-dependent) two-photon drive $\tilde{\lambda} [\omega]$. While the frequency-dependence of both $\Sigma_d [\omega]$ and $\tilde{\lambda} [\omega]$ prevents any attempt at directly mapping our system onto a DPA, on-resonance this resemblance provides a very direct way to understand the emergence of the negativity of the spectral function $A[\omega]$ in our system.

For a true resonant DPA, applying the results of Subsec.~\ref{subsec:resonant_paramps} yields the spectral function
\begin{equation}	\label{eq:A_DPA}
	A^{(\mathrm{DPA})} [\omega] = \frac{\kappa \left( \kappa^2/4 - | \Lambda |^2 + \omega^2 \right)}										{\left( \kappa^2/4 - |\Lambda|^2 - \omega^2 \right)^2 + \kappa^2 \omega^2}. 
\end{equation}
As mentioned, we cannot directly map our system onto a DPA. However, \textit{on-resonance}, the effective susceptibility matrix resulting from the effective cavity dynamics Eqs.~\eqref{eq:cavity_EOM} and \eqref{eq:EOM_terms} looks \textit{exactly} like that of a DPA but with effective damping $\kappa_{\mathrm{eff}} [0] = \kappa - 2 \mathrm{Im} \,\Sigma_d [0]$ and effective parametric drive $\tilde{\lambda}[0]$. We can therefore find $A[0]$ for our system by directly substituting $\kappa \rightarrow \kappa_{\mathrm{eff}} [0] = \kappa - 2 \mathrm{Im} \, \Sigma_d [0]$ and $\Lambda \rightarrow \tilde{\lambda} [0]$ into Eq.~\eqref{eq:A_DPA}.

One finds that the effective damping becomes negative ($\kappa_{\mathrm{eff}} [0] < 0$) when
\begin{equation}
	\frac{2 \lambda}{\gamma} > \sqrt{1+C_0},
\end{equation}
which is precisely where the spectral function $A[0]$ becomes negative. This is, of course, no coincidence: in this regime, the on-resonance ($\omega = 0$) effective susceptibility of our system looks exactly like that of an \textit{unstable} DPA. This provides an intuitive understanding of the origin of the negative spectral function and associated negative effective photon temperature in our system.

It is important to recall that in reality, the system remains stable until $2 \lambda / \gamma > 1 + C_0$. Of course, there is no contradiction here; the stability of the system depends on the location of the \textit{poles} of the susceptibility (Green's function), not on the value of the susceptibility at any particular frequency.

\subsection{The detuned DPA}
In Subsec.~\ref{subsec:resonant_paramps}, we showed how a resonant DPA has a spectral function which is positive at all frequencies. This is not in general the case for a \textit{detuned} DPA, where the pump field is applied at a frequency $\omega_p \ne 2 \omega_c$. In a frame rotating at $\omega_p / 2$ to make the paramp Hamiltonian time-independent, the detuning shows up as a photon energy $- \Delta \hat{a}^{\dagger} \hat{a}$,
\begin{equation}
	\hat{H} = - \Delta \hat{a}^{\dagger} \hat{a} + \frac{i}{2} \left( \Lambda \hat{a}^{\dagger} \hat{a}^{\dagger} - \Lambda^* \hat{a} \hat{a} \right),
\end{equation}
where $\Delta \equiv \omega_p / 2 - \omega_c$. In such a system, the stability regime is extended from $ |\Lambda|^2 < \kappa^2/4$ to $ | \Lambda|^2  < \kappa^2 / 4 + \Delta^2$. In the extension of the stability regime where $\kappa^2 / 4 < | \Lambda |^2 < \kappa^2  /4 + \Delta^2$, there are frequencies at which the cavity spectral function becomes negative.

We stress that this negativity occurs for different reasons than in our system. In the preceding Subsec.~\ref{subsec:negative_A_mapping}, we showed how $A [\omega] < 0$ emerges in our system as a result of a negative total frequency-dependent damping $\kappa_{\mathrm{eff}} [\omega] = \kappa - 2 \mathrm{Im} \, \Sigma_d [\omega]$. This is completely different from the detuned DPA, where the matrix self-energy $\matr{\Sigma} \equiv i (\matr{\chi}_0^{-1} - \matr{\chi}^{-1})$ is purely off-diagonal (and hence where $\mathrm{Im} \, \Sigma_d = \mathrm{Im} \, 0 = 0$).
%%%%%%%%%%%%%%%%%%%%%
\section{Connection between spectral function and probe-field reflection}	\label{sec:spectral_function_appendix}

In this section, we derive the connection between the cavity spectral function $A[\omega]$ and the power reflection coefficient describing
the reflection of probe signals incident on the cavity through a weakly coupled auxiliary waveguide (as given in Eq.~(\ref{eq:spectral_function_reflection})).
%In the main text we have discussed the possibility for the cavity spectral function $A[\omega]$ to become negative. Here we spend a moment fleshing out the meaning of this negativity. For a Hamiltonian, time-independent system, with a density matrix diagonal in the energy basis, $A[\omega] < 0$ indicates a population inversion between states separated in energy by $\hbar \omega$. \BL{As discussed in the main text, such} considerations provide little in the way of interpretive insight, and so we consider a more operational approach.
We couple the cavity to a second input-output reservoir (input modes $\hat{c}_{\mathrm{in}}(t)$) at a rate $\kappa^{\prime} \ll \kappa$.  Combining the standard input-output boundary condition with linear response theory, we find that the average output field in this auxiliary waveguide is given by:
\begin{equation}
	\langle \hat{c}_{\mathrm{out}} [\omega] \rangle = \left(1  -i \kappa^{\prime} G^R [\omega] \right) \langle \hat{c}_{\mathrm{in}} [\omega] \rangle
		- \kappa^{\prime} \chi_{d d^{\dagger}} [\omega] \langle \hat{c}_{\mathrm{in}} [-\omega] \rangle^{*}.
\end{equation}

Here, $G^R$ is the cavity retarded Green's function as defined in the main text, and $\chi_{d d^{\dagger}}$ is the off-diagonal component of the cavity susceptibility matrix expressed in the field operator basis $(\hat{d}, \hat{d}^{\dagger})^T$.
By detuning the probe from the cavity resonance (i.e.\@ $\omega = \epsilon \ne 0$ in the rotating frame, $\omega = \omega_c + \epsilon$ in the lab frame), we can have $\langle \hat{c}_{\mathrm{in}} [-\omega] \rangle = 0$ and we eliminate the term involving the anomalous susceptibility $\chi_{d d^{\dagger}}$. The amplitude reflection coefficient is then $(1 - i \kappa^{\prime} G^R [\omega] )$, and taking the magnitude-square to get the power reflection coefficient, we find
\begin{equation}
	\mathcal{R} = 1 - \kappa^{\prime} A[\omega] + \mathcal{O} \left( (\kappa^{\prime})^2 \right)
\end{equation}
as stated in the main text. As is also mentioned in the main text, this result also holds on-resonance, if one averages over the phase of the incident  drive in the auxiliary waveguide; in this case, the contribution $\propto \chi_{d d^{\dagger}}$ in $\langle \hat{c}_{\mathrm{out}} [\omega] \rangle$ averages away.  

%%%%%%%%%%%%%%%%%%%%%%%
\bibliography{optomechanics,amplification,squeezing_applications}

%\end{widetext}
\end{document}